\documentclass[runningheads]{llncs}
\usepackage[T1]{fontenc}
\usepackage{subcaption}
\usepackage{booktabs}
\usepackage{enumitem}

\usepackage{float} 
\usepackage{pifont}
\usepackage{graphicx}
\usepackage[table]{xcolor}
\usepackage{todonotes}
\usepackage{multirow}
\usepackage{siunitx}
\usepackage{hyperref}   
\usepackage[ruled,vlined]{algorithm2e}
\usepackage{listings}
\usepackage{colortbl}
\usepackage{amssymb}
\usepackage{verbatim}
\usepackage{lscape}

\begin{document}
\newcommand\mycommfont[1]{\footnotesize\ttfamily\textcolor{black}{#1}}
\SetCommentSty{mycommfont}

\SetKwInput{KwInput}{Input}                
\SetKwInput{KwOutput}{Output}              
\title{How to Diversify \texttt{any} Personalized Recommender?}
\titlerunning{A User-centric Pre-processing approach}
%
\author{Manel Slokom\orcidID{0000-0002-9048-1906} \and 
Savvina Daniil\orcidID{0000-0001-8888-2869} \and
Laura Hollink\orcidID{0000-0002-6865-0021}}

\authorrunning{M. Slokom}

\institute{Centrum Wiskunde \& Informatica, Amsterdam, The Netherlands \email{manel.slokom@cwi.nl}, \email{Savvina.Daniil@cwi.nl}, \email{laura.hollink@cwi.nl}} 

\maketitle              

\begin{abstract}

In this paper, we introduce a novel approach to improve the diversity of Top-N recommendations while maintaining accuracy. 
Our approach employs a user-centric pre-processing strategy aimed at exposing users to a wide array of content categories and topics.
We personalize this strategy by selectively adding and removing a percentage of interactions from user profiles. 
This personalization ensures we remain closely aligned with user preferences while gradually introducing distribution shifts. 
Our pre-processing technique offers flexibility and can seamlessly integrate into any recommender architecture.
We run extensive experiments on two publicly available data sets for news and book recommendations to evaluate our approach.
We test various standard and neural network-based recommender system algorithms.
Our results show that our approach generates diverse recommendations, ensuring users are exposed to a wider range of items. 
Furthermore, using pre-processed data for training leads to recommender systems achieving performance levels comparable to, and in some cases, better than those trained on original, unmodified data.
Additionally, our approach promotes provider fairness by facilitating exposure to minority categories.\footnote{Our GitHub code is available at: \url{https://github.com/SlokomManel/How-to-Diversify-any-Personalized-Recommender-}}


\end{abstract}

\section{Introduction}
\label{sec:intro}
In today's digital age, personalized recommender systems form a solution to the information overload problem. 
They filter the enormous amount of information available online to only the top-N items - e.g. news articles, or books - that are most relevant to a user~\cite{liu2019diverse,meng2023survey}.
While personalized recommendations ensure relevance to users' interests, excessive personalization risks closing users within filter bubbles~\cite{pariser2011filter,Michiels2023How}. 
Filter bubbles are environments created by personalization algorithms, where users are exposed solely to familiar information or opinions.
This can potentially harm the democratic process and exacerbate polarization~\cite{pariser2011filter}.

To address this, diversity in recommender systems has become an active research area. 
This trend aligns with a larger movement in the community towards evaluating not only accuracy but also other metrics, which we refer to as `beyond accuracy', such as diversity, privacy, and fairness~\cite{Kaminskas2016Diversity,zhao2023fairness,Yao2023Fairness}.
Although diversity and provider fairness are rooted in distinct normative considerations~\cite{Ekstrand2022}, diversity aims to present a wide range of items, while provider fairness ensures equitable exposure for each content source~\cite{Liu2024Measuring}, in our case increasing diversity also improves provider fairness by enhancing the visibility of items from minority providers.
Several approaches have emerged to improve beyond-accuracy metrics. 
In a common categorization, we distinguish between three types of approaches~\cite{zhao2023fairness}: (i) pre-processing, (ii) in-processing, and (iii) post-processing. 
Pre-processing approaches involve modifying input data - such as user profiles - before the recommendation process. 
Several works use pre-processing to improve fairness~\cite{Rastegarpanah2019Fighting,pmlr-v81-ekstrand18a,boratto2021interplay,Boratto2024FairGNN}. 
\cite{slokom2021PerBlur} propose pre-processing to ensure privacy. 
In-processing approaches build beyond accuracy metrics directly into the recommendation algorithm itself, e.g. through multi-objective optimization
~\cite{Zheng2021DGCN,Stamenkovic2022Choosing,raza2022accuracy,oliveira2017multiobjective,Hansen2021Shifting,wu2022end}.  
Post-processing approaches such as re-ranking have been used for both diversity~\cite{Mansoury2020FairMatch} and fairness~\cite{Zehlike2017FA*IR,Zehlike2022FAIR}. 
We notice that the majority of research into diversity has focused on in-processing and post-processing approaches; relatively little is known about the use of pre-processing to increase diversity. 

In this paper, we fill this gap and present a pre-processing approach to diversify the output of a recommender system. 
We alter user profiles that are input to the system to achieve greater diversity in the recommender's output. 
Pre-processing is interesting in this respect, as such an approach can be combined with a wide range of algorithms. 
Any personalized recommender architecture takes a user profile as input, and can thus be used in combination with our approach to diversification. 
In comparison, in-processing approaches are bound to a specific algorithm~\cite{zhao2023fairness}. 
In addition, pre-processing allows diversification beyond what would have been recommended by the initial approach. 
In contrast, the effect of post-processing is independent of the model and may be constrained by the set of items the algorithm can recommend. 
Our approach is user-centric, focusing on the personalized addition of items to user profiles. 
This is because the input data for recommender systems is highly personal. 
A user profile is more than a list of past interactions with items; it encodes parts of the user's preferences, abilities, and characteristics. 
Altering a user profile risks diluting this rich encoding of the user's taste. 
By adopting a personalized approach, we mitigate this risk, maintaining accuracy without compromising the integrity of the user profile. 
We opt to alter the user profile directly instead of altering processed versions of it, e.g. a latent representation. 
This increases the potential for transparency and explainability, allowing users to review their (altered) profiles.

Our main research question focuses on how to alter user profiles to improve the diversity of top-N recommendations without compromising recommendation accuracy.
We present two variants of our approach: one involves adding interactions to user profiles, while the other includes both adding and removing interactions to remain close to the original profile size. 
We explore different levels of manipulations ranging from 1\% to 10\%. 
We perform extensive evaluations to test which variant achieves optimal accuracy and diversity. 
We evaluate the effectiveness of our approach on two public data sets: 
MIND for news recommendation, where we aim to diversify news categories, and GoodBook for book recommendation, where we aim to diversify genres.
As mentioned above, our approach can be integrated with any personalized recommender system. 
In this study, we test combinations with seven algorithms ranging from collaborative filtering (CF) to neural network recommenders.

We report accuracy (MRR, nDCG), normative diversity (calibration via divergence), descriptive diversity (coverage, Gini index), and provider fairness (fair-nDCG).
Our findings indicate that using pre-processed data for training can lead to recommender systems achieving comparable or improved accuracy compared to those trained on original data. 
Regarding diversity, calibration metrics consistently improved for certain algorithms, while coverage and the Gini index showed mixed results across different levels of pre-processing. 
Additionally, pre-processed data consistently resulted in higher fair-nDCG scores, indicating enhanced exposure fairness and better representation of minority categories.

\section{Background and Related Work}

This section, first, provides an overview of existing work on pre-processing approaches to increase beyond-accuracy factors. 
Then, we examine existing measures used to evaluate diversity in recommender systems.

\subsection{Pre-processing approaches}
Pre-processing approaches involve modifying input data before feeding it into the recommender system~\cite{zhao2023fairness}. 
These approaches offer flexibility, as they can be integrated with any recommendation algorithm without altering its core design~\cite{meng2023survey,Boratto2024FairGNN}. 
Inspired by data poisoning attacks, \cite{Rastegarpanah2019Fighting} introduces ``antidote'' data by simulating new users and ratings in the training set to improve consumer fairness in matrix factorization models. 
Similarly, \cite{slokom2021PerBlur,weinsberg2012blurme} proposes a gender obfuscation method to protect user privacy while preserving recommendation quality and reducing gender stereotypes.
Balancing user groups through re-sampling is another common technique. 
\cite{pmlr-v81-ekstrand18a} creates gender-balanced training data by adjusting the user distribution. 
While this reduces recommendation accuracy slightly, it mitigates gender disparities in data sets like Movielens. 
\cite{boratto2021interplay} explores provider fairness by up-sampling interactions favoring minority groups and shows improvements in coverage and reduced disparate impacts without sacrificing accuracy. 
Our study builds on this line of work by proposing a personalized pre-processing method that selectively extends user profiles 
aiming to enhance diversity without compromising recommendation accuracy.

\subsection{Measuring diversity}
Diversity in recommender systems was initially defined as the opposite of similarity, i.e., \( 1 - \text{similarity} \)~\cite{Smyth2001Similarity}. 
In~\cite{Ziegler2005Improving}, Intra-List Similarity (ILS) was introduced to measure the average diversity within a recommendation list, independent of the item order. 
Other metrics include expected intra-list diversity, which is sensitive to item ranking, and aggregate diversity, which measures the diversity of items across all user recommendations~\cite{Vargas2011Rank}, and the Gini index quantifying distributional inequality~\cite{Daniel2009Blockbuster}. 
Item coverage is also used as a descriptive measure of diversity, calculated as the proportion of unique items recommended across users relative to the entire catalog.
While traditional descriptive diversity metrics provide useful insights, they only partially address the broader challenges in recommender systems~\cite{Treuillier2022Being}. 
Recent research has extended the focus to normative diversity, emphasizing exposure diversity based on principles tailored for specific domains. 
In~\cite{Helberger2018Exposure}, the authors outline principles for achieving exposure diversity in recommendations, and \cite{Vrijenhoek2022RADio} introduces RADio, a rank-aware divergence framework designed to evaluate diversity according to normative goals.
In this study, we use both descriptive and normative metrics.

\section{User-Centric Data Pre-processing}
\label{sec:method}

\begin{figure}[!htb]
    \centering
    \includegraphics[width=0.6\paperwidth]{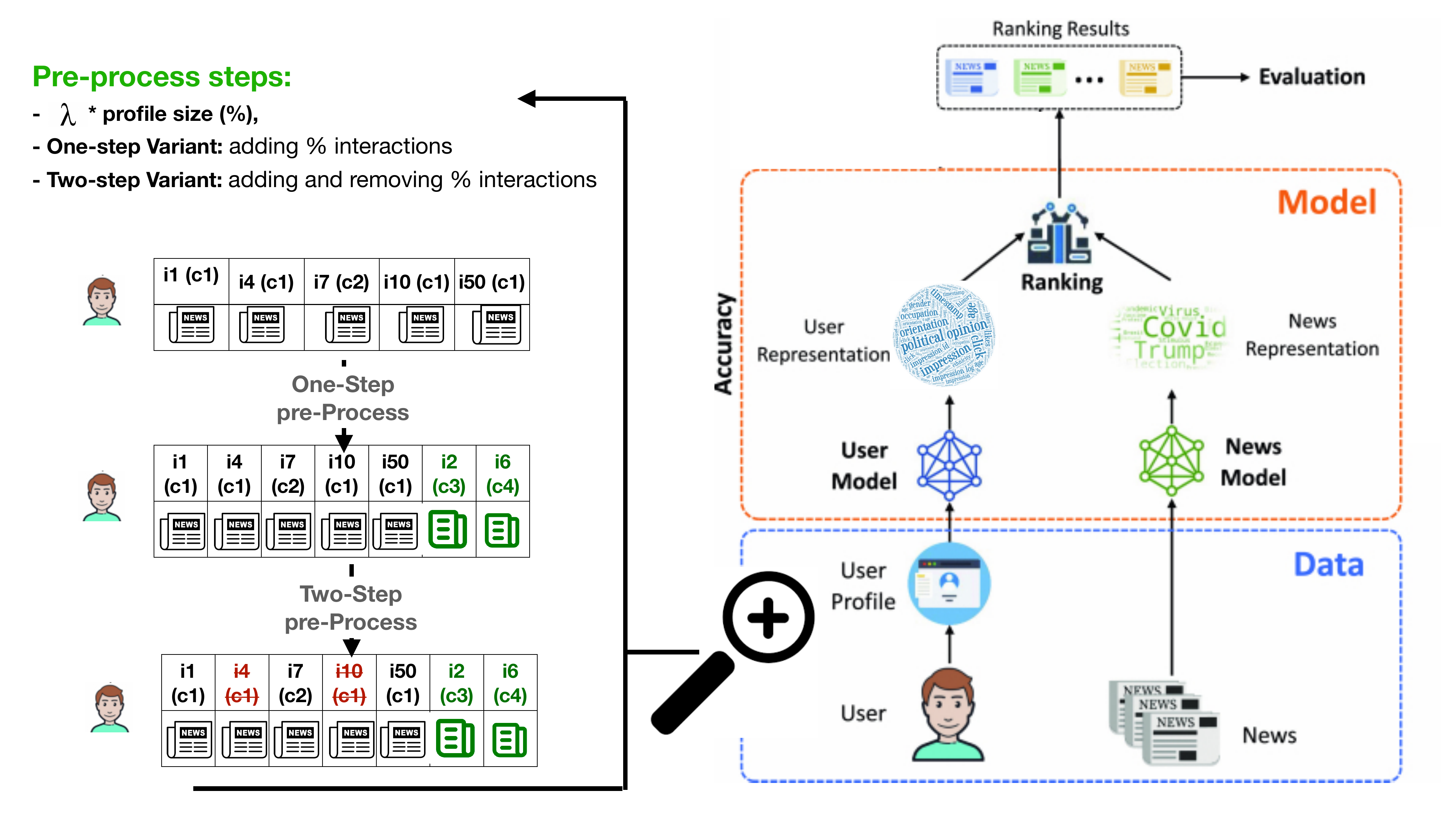}
    \caption{Our pre-processing approach and its role in a recommendation architecture. \( \{i1, i4, i7, i10, i50\} \) are items in user \(u\)'s past interactions and \( \{c1, c2, c3, c4\} \) are categories of these items. In the one-step variant, personalized items \(i2\) and \(i6\) are added (green). In the two-step variant, items \(i4\) and \(i10\) are removed (red).}
\label{fig:diagram-fair}
\end{figure}

The objective of our personalized pre-processing approach is to increase the recommendation diversity 
while maintaining the accuracy. 
For this purpose, we alter user profiles through strategic addition and removal of interactions. 
Our approach is founded on two principles: (1) that items added to the user profile should be sufficiently different from what a user has interacted with before, and (2) that those items should closely align with the user's preferences.   Figure~\ref{fig:diagram-fair} illustrates our approach. 
Three key inputs drive this process. 
First, each item in the system is associated with one or more categories (\(Item\_categories\)). 
Second, we use logistic regression to identify the categories most representative of each user, based on their past interactions (\(User\_categories\)). 
Lastly, we apply userKNN to generate an initial list of items that might be relevant for each user (\(Initial\_predictions\)).
Based on these three inputs, we generate a personalized list of items that can be added to the user's profile ($ Personalized\_items\_for\_addition$). 
For this list, we select items from the user's $Initial\_predictions$ that are associated with categories not in their $User\_categories$.
Section~\ref{sub:Pers_user} further explains how we arrive at personalized lists of items for addition.  

Our approach has two variants. 
The one-step variant adds interactions to a user's profile. 
The amount of added items is determined by a parameter $\lambda$. 
The two-step variant, on top of that, also removes interactions from a user's profile.
After adding items as in the one-step approach, this variant randomly removes items that were not recently added. 
This ensures that the profile size remains approximately equal to its original size. 
Section~\ref{sub:One} describes the two variants. 

\subsection{Creating personalized lists of items for addition}
\label{sub:Pers_user}
Here, we explain how we create $User\_categories$, $Initial\_predictions$, and based on that, $Personalized\_items\_for\_addition$. 

\paragraph{User\_categories}
For each user, we create a \textit{User\_categories} list that contains categories that are representative of that user. 
For this purpose, we train a 
logistic regression model on labeled training data. 
We choose logistic regression for its simplicity and interpretability, allowing us to easily understand the relationship between user profiles and categories.
Additionally, its coefficients help identify only the truly associated categories, preventing overgeneralization.
Lastly, logistic regression has shown strong performance on similar tasks in prior research~\cite{weinsberg2012blurme,slokom2021PerBlur}.
The input for the model is the initial user-item matrix \( \mathcal{D} \), and the target is whether an item belongs to a specific category. 
We transpose the input data \( \mathcal{D} \) into an item-user matrix \( X_{iu} \), which indicates whether an item (in rows) \( i \) has interacted with user \( u \).
We apply one-hot encoding to the target feature, i.e., categories. 
Then, we have binary category labels \( \mathbf{y}^c \), such that for each category \( c \), a binary vector \( \mathbf{y}^c \), where \( y_i^c = 1 \) if item \( i \) belongs to category \( c \), and \( y_i^c = 0 \) otherwise.
For each category \( c \), we train a logistic regression:
\small{
\[
P(y_i^c = 1 \mid X_{i1}, X_{i2}, \ldots, X_{iM}) = \frac{1}{1 + e^{-(\beta_0^c + \beta_1^c X_{i1} + \beta_2^c X_{i2} + \ldots + \beta_M^c X_{iM})}}
\]}

Where:
\( y_i^c \) is the binary label indicating if item \( i \) belongs to category \( c \).
\( X_{iu} \) is the interaction of item \( i \) with user \( u \).
\( \beta_0^c \) is the intercept term for category \( c \).
\( \beta_u^c \) is the coefficient for the interaction of user \( u \) with items in category \( c \).
The \( \beta_u^c \) coefficients capture the extent to which users are correlated with categories. 
Users with higher positive coefficients are more indicative of the category; 
we select user-category pairs with coefficients above 0 for the $User\_categories$ lists. 

\paragraph{Initial\_predictions} 
The goal here is to predict a list of items that a user might be interested in interacting with, based on the initial user-item matrix \( \mathcal{D} \). 
Any recommender system could be used for this step. 
We use an imputation approach to fill in the missing values in the user-item matrix~\cite{lakshminarayan1999imputation}. 
The goal of the imputation is to infer missing values in the data in such a way that improves the overall performance of recommender systems trained on that data~\cite{su2008imputed,slokom2021PerBlur}. 
Specifically, we impute to derive a confidence score that allows us to choose the items that will be added to the profile. 
For simplicity, we opt for a userKNN-based approach, as implemented in~\cite{fancyimpute}. 
We use the number of neighbors that are used for each prediction as a relevance score. 
This enables us to account for the fact that some users may have more neighbors available for some items than others. 
We fix a cutoff $k$, 
to ensure that only the most relevant items are added to the list of initial predictions for each user. 
Algorithm~\ref{algo:perso} (first part) summarizes this process. 

\begin{algorithm}[!htb]
\tiny{
\DontPrintSemicolon
  
  \KwInput{}
  \begin{itemize}
      \item User-item matrix \( \mathcal{D} \) (\( \mathbb{N} \) users, \( \mathbb{M} \) items),
      \item Representative categories for user \texttt{u} ($User\_categories^{u}$), 
      \item List of items per category ($Item\_categories$)
      
  \end{itemize}
  
  \KwOutput{$Personalized\_items\_for\_addition^{u}$} 
  
  \tcp*{0. $Initial\_Predictions^{u}$: Generate predictions using UserKNN.}
  
  \For {(user \( \texttt{u} \in \mathbb{N} \))}{
    \For{(item \( \texttt{i} \in \mathbb{M} \))}{
        Similarity computation finds nearest neighbor candidates;
        
        Sort predicted items based on the number of possessed neighbor candidates;
        
        $Initial\_Predictions^{u}$ contains the top-k predicted items for user $u$
    }
  }

  \tcp*{1. \( Personalized\_items\_for\_addition^{u}\): Create a personalized list of items that can be added to the profile of user \( u \).
  }
  
  \For {(user \( \texttt{u} \in \mathbb{N} \))}{
  
      \For {item \( i \in Initial\_Predictions^{u} \)}{ 
        \If{(category of item \( i \notin User\_categories^{u} \))}{
          \( Personalized\_items\_for\_addition^{u} = Personalized\_items\_for\_addition^{u}\).add(\( i \)) 
        }
      }     
  }
\caption{Personalized list of indicative items per user.}
\label{algo:perso}
}
\end{algorithm}

\paragraph{Personalized\_items\_for\_addition}
For each user {u}, we create a list of items that can be added to their profile. 
This involves selecting items that are present in \( Initial\_Predictions^{u} \) and that are associated with a category that is not in \( User\_categories^{u} \). 
In other words, we select items that are probably relevant to the user and that are from a category with which they have not had much interaction. 
As can be seen in Algorithm~\ref{algo:perso} (second part), items in this list are ranked based on the ranking in \( Initial\_Predictions^{u} \).

\subsection{One-step and Two-step pre-processing}
\label{sub:One}
Algorithm~\ref{algo:1step} presents the basic skeleton of our one and two-step pre-processing.
For each user, we calculate the number of items to be added based on their initial profile size and parameter \( \lambda \), which represents the desired addition level as a percentage of the initial profile.
We then iteratively pick items from the personalized list 
\( Personalized\_items\_for\_addition^{u} \) 
and add them to the user's profile.
\begin{algorithm}[h]
\tiny{
\DontPrintSemicolon
  
  \KwInput{}
  \begin{itemize}
      \item \( \lambda \): level of pre-processing
      \item User-item matrix \( \mathcal{D} \) (\( \mathbb{N} \) users, \( \mathbb{M} \) items)
      \item \( \mathcal{D}' \) (\( \mathbb{N} \) users, \( \mathbb{M} \) items) is a copy of \( \mathcal{D} \) at time $t=0$
      \item \( \mathcal{D}'' \) (\( \mathbb{N} \) users, \( \mathbb{M} \) items) is a copy of \( \mathcal{D}' \) after the one-step pre-processing. 
      \item Initial count: user profile size at time \( t = 0 \) 
      \item Interaction count: user profile size after adding $\lambda \%$ extra interactions
      \item Removal threshold
      \item Total added = 0 at time \( t = 0 \) 
  \end{itemize}
  
  \KwOutput{One-step and two-step pre-processed data \( \mathcal{D^{'}} \) and \( \mathcal{D^{''}} \) (\( \mathbb{N} \) users, \( \mathbb{M} \) items)}
  
  \tcp{1. One-step: adding extra items}
  
  \For{(user \( u \) in \( \mathbb{N} \))}{
    count = initial count [u] \( \ast \lambda \)
    
    added = 0
    
    \While{added \( < \) count}{
        i = pick the item in the highest position in \( Personalized\_items\_for\_addition^{u} \)
        
        \If{\( \mathcal{D}' [u, i] == 0 \)}{
            \( \mathcal{D}' [u, i] = 1 \)
            
            added += 1
        }
    }
    }
    Total added += added

   \tcp{2. Two-step pre-processing: Removing certain items}

    \For{user $ u $ in $\mathcal{N} $}{
        \If{Interaction count $ \geq $ Removal threshold}{ 
            removed count += 1
  }
}

To be removed = Total added/removed count 

\For{user $ u $ in $\mathcal{N} $}{
  \If{Interaction count $ \geq $ Removal threshold}{
    removed = 0
    
    \While{removed $ < $ To be removed}{
      i = Randomly pick an item that has not recently been added 
      
      \If{$\mathcal{D}'[u, i] \neq 0$}{
        $\mathcal{D}''[u, i] = 0$ \\
        removed += 1
      }
    }
  }
}
  
\caption{One and two-step pre-processing algorithm.}
\label{algo:1step}
}
\end{algorithm}
The two-step pre-processing takes the data generated by one-step pre-processing as input and focuses on removing 
interactions from a user's profile. 
The removal process helps in maintaining the density of the modified data close to that of the original data. 
The removal is performed in such a way that the total number of user interactions for each item remains proximate to the total in the original data set. 
We distribute the removal of interactions evenly across all users.
To account for users with very short profiles, we set a threshold of 20 that exempts them from having interactions removed.

The removal step is driven by several motivations. 
First, it respects the nature of recommender system data, which is typically sparse and follows a long-tail distribution. 
Maintaining these properties, we avoid unintended shifts that might impact our results~\cite{deldjoo2021explaining}. 
Second, it addresses the potential impact of the increased data density from the addition step. 
It has been shown in~\cite{pmlr-v81-ekstrand18b} that the performance of recommendation algorithms is not necessarily and solely impacted by the profile size of the users. 
When we adjust for the added interactions, we aim to ensure that observed changes in performance are due to our pre-processing approaches rather than shifts in the data's size.
Finally, we apply the removal selectively to long profiles, ensuring minimal disruption to users with shorter profiles, thus maintaining a fair representation across the data.

\section{Experimental Setup}
In this section, we start by providing an overview of our data sets and variants. 
Next, we describe recommender system algorithms used for both news and book recommendations in our experiments.

\paragraph{Data sets.}
We test our approach on two publicly available data sets: the MIND data~\cite{wu2020mind} for news recommendations and the GoodBook data for book recommendations. 
Statistics for the data sets can be found in Table~\ref{tab:datasets}. 
From MIND, we use a subset of the published train data and the entire published test data. 
In the GoodBook data, we transform explicit ratings into implicit, considering ratings with a rating $>= 3$ as implicit indicators of preference. 
We randomly sample 80\% of each user profile for the training set and we keep the remaining 20\% for the test set.
It is important to note that the choice of static splitting plays a key role in preventing the data pre-processing from adding items to the test set. 
Our pre-process is solely applied to the training set, with careful consideration to avoid adding interactions already present in the test set. 
As a result, the test set remains invariant across all experimental conditions. 
It is worth noting that the added items through pre-processing are not recommended anymore. 
While these added items could still be of interest, this is not a concern for MIND, as the added items are older and less likely to be relevant. 
However, for GoodBook, where random splitting is used, this could pose a challenge since the added items may be more recent and potentially impactful for recommendations. 
We recommend the use of temporal splitting whenever possible to ensure that recent items are recommended.
In the MIND data, there are 17 categories, whereas in the GoodBook data, there are 31 genres. 
A news article is assigned to a single category, while a book is associated with multiple genres. 
Both news categories and book genres are considered as \textit{categories} in our approach as discussed above. 
\begin{table}[htb]
\centering
\caption{Statistics of the data sets used in the experiments.}
\label{tab:datasets}
\scriptsize{
\begin{tabular}{lcccccc}
\toprule
\textbf{Data set}     & \textbf{Type}   & \textbf{\#Users} & \textbf{\#Items} & \textbf{\#Clicks} & \textbf{Sparsity (\%)} & \textbf{\#Categories} \\ \midrule
\multirow{2}{*}{MIND News}   & Training & 1,000  & 26,740 & 9,368  & 99.58  & 17 \\
                             & Testing  & 5,000  & 18,723 & 15,557 & 99.82  & 16 \\ 
\multirow{2}{*}{GoodBook}    & Training & 943    & 729    & 8,477  & 98.77  & 31 \\
                             & Testing  & 943    & 688    & 4,715  & 99.27  & 31 \\ 
\bottomrule
\end{tabular}}
\end{table}

\paragraph{Overview of data variants and experimental parameters.} 
In our experimental setup, we generate several data variants to evaluate the impact of our pre-processing approach. We begin with the original data and apply both one-step and two-step pre-processing. For each approach, we create four variants using different values of the parameter $\lambda $. 
Additionally, we include a baseline where $\lambda$ interactions are added and removed randomly.\footnote{Further details about our pre-processed data and results of our baseline are available in the supplementary material due to space constraints \url{https://surfdrive.surf.nl/files/index.php/s/QaUVGB4Vh3sv7kB}}
Lastly, to ensure that the relevant items are adequately included within the two data sets, we set the maximum cutoff value $k$ differently when we use userKNN to create the lists of \(Initial\_predictions\); $k$ is set to 100 neighbors for the GoodBook data, and to 50 neighbors for the MIND data.

\paragraph{Recommender system algorithms.}
\label{sub:recsysAlgo}
We select several recommendation algorithms, ranging from standard collaborative filtering to advanced neural-based models. 
This demonstrates the adaptability of our approach across various recommendation architectures.
It also allows us to test our approach with the different algorithms that are currently most prevalent in the two domains that we work with, news and books.  

For news recommendations, we use three state-of-the-art neural network recommendation algorithms:
\textit{NRMS} uses multi-head self-attention networks~\cite{wu2019NRMS}. 
\textit{NPA} uses personalized attention networks to learn news and user representations~\cite{Chuhan2019NPA}. 
\textit{LSTUR} is a news recommender focusing on users' long and short-term preferences~\cite{Fangzhao2019LSTUR}. 
To account for the randomness in the algorithms, we repeat each experiment 5 times and show average performance.
We use the recommenders package from Microsoft.\footnote{\url{https://github.com/recommenders-team/recommenders}}
For hyper-parameter tuning, we set epochs to 50. 
We follow the parameters as suggested by the recommenders package.
For book recommendations, we use CF algorithms:
    \textit{MostPop} is a non-personalized algorithm recommending the most popular items. 
    \textit{ItemKNN} is the item-based K-nearest neighbor algorithm~\cite{Deshpande2004Item}. 
    \textit{ImplicitMF} is matrix factorization trained with alternating least squares~\cite{Yifan2008ImplicitMF}. 
    \textit{BPR} is a matrix factorization algorithm using Bayesian personalized ranking~\cite{Rendle2009BPR}. 
We use the Lenskit toolkit for our implementation of these algorithms.\footnote{\url{https://github.com/lenskit/lkpy/tree/main}}
We perform hyperparameter tuning on the training set to optimize the performance. 
We use the AllItems strategy for our candidate item selection, as proposed in~\cite{Bellogin2011Precision}. 
We measure Normalized Discounted Cumulative Gain (nDCG) and Mean Reciprocal Rank (MRR).
For the MIND data, we report MRR for the entire recommendation list as implemented in Recommenders, whereas, for the GoodBook data, we measure MRR@10 as is common with Lenskit.

\paragraph{Diversity and fairness metrics.} 
We apply a normative diversity metric, calibration, which assesses how closely recommendations align with user preferences inferred from their past interactions. 
We quantify calibration as the difference between the distributions of categories in user preferences and recommendations, measured with the Kullback-Leibler (KL) and Jensen-Shannon (JS) divergence metrics, as described in~\cite{Vrijenhoek2022RADio}.  
We also compute two descriptive diversity metrics: item coverage (coverage) and the Gini index (gini). 
Item coverage represents the proportion of unique items recommended across users relative to the entire catalog, while the Gini index captures inequality in item distribution among users. 
To facilitate interpretation, we report 1 $-$ Gini.
For all three diversity metrics, higher values indicate greater diversity. 

Finally, we apply discounted cumulative fairness, or fair-nDCG~\cite{Pitoura2021Fairness}. 
Fair-nDCG is an adaptation of nDCG that incorporates fairness by emphasizing exposure to underrepresented categories. 
It accumulates the gain for items belonging to the protected group while applying a logarithmic discount based on their rank, emphasizing their presence at higher ranks to ensure fair representation in the recommendation list. 
In our experiments, we designate the least-represented categories in the data set as the protected group. 
For MIND, these categories are: kids, weather, video, music, autos, movies, middleeast, and northamerica. For GoodBook, they include: music, poetry, horror, spirituality, sports, christian, comics, manga, cookbooks, psychology, and art.

\section{Recommendation Performance}

In Table~\ref{tab:Step-preprocess-accDiv} and Table~\ref{tab:Step-preprocess-GoodRead_AccDiv}, we present our recommendation results using one-step and two-step pre-processing techniques on the MIND and GoodBook data sets, respectively. 
We measure the accuracy of the recommendation using MRR, nDCG@5, and nDCG@10. 
\begin{table}[htb]
\caption{Recommendation performance for the MIND data. 
Recommendations with the highest scores are marked in bold, comparing original and pre-processed data. Statistical significance is indicated by a $^*$, based on paired t-tests comparing $\lambda$ values to the baseline ($\lambda = 0$). Standard deviation is denoted by $\pm$.}
\label{tab:Step-preprocess-accDiv}
\scriptsize{
\resizebox{\textwidth}{!}{%

\begin{tabular}{cccccccccccccc}
\cline{3-14}
\multicolumn{2}{c}{\multirow{2}{*}{\textit{\textbf{}}}}   & \multicolumn{6}{c}{One-Step Pre-process}                                                                                                                                                                                                                                        & \multicolumn{6}{c}{Two-Step Pre-process}                                                                                                                                                                                                                                \\ \cline{3-14} 
\multicolumn{2}{c}{}                                      & \multicolumn{3}{c}{Accuracy}                                                                                                                                                                                    & \multicolumn{2}{c}{Calibration @10}         & Descriptive @10 & \multicolumn{3}{c}{Accuracy}                                                                                                                                                                                    & \multicolumn{2}{c}{Calibration @10} & Descriptive @10 \\ \hline
\textit{\textbf{Algorithms}}    & \textit{$\lambda$ (\%)} & \textit{MRR}                                                        & \textit{nDCG@5}                                                     & \textit{nDCG@10}                                                    & \textit{KL}              & \textit{JS}      & \textit{Cov}    & \textit{MRR}                                                        & \textit{nDCG@5}                                                     & \textit{nDCG@10}                                                    & \textit{KL}      & \textit{JS}      & \textit{Cov}    \\ \hline
\multirow{5}{*}{\textbf{NRMS}}  & \textit{0}              & \begin{tabular}[c]{@{}c@{}}0.2688\\ $\pm$ 0.001\end{tabular}           & \begin{tabular}[c]{@{}c@{}}0.2914\\ $\pm$ 0.002\end{tabular}           & \begin{tabular}[c]{@{}c@{}}0.3592\\ $\pm$ 0.001\end{tabular}           & 2.1947                   & 0.3918           & 1342            & \begin{tabular}[c]{@{}c@{}}0.2688\\ $\pm$ 0.001\end{tabular}           & \begin{tabular}[c]{@{}c@{}}0.2914\\ $\pm$ 0.002\end{tabular}           & \begin{tabular}[c]{@{}c@{}}0.3592\\ $\pm$ 0.001\end{tabular}           & 2.1947           & 0.3918           & 1342            \\
                                & \textit{1}              & \begin{tabular}[c]{@{}c@{}}0.2686\\ $\pm$ 0.002\end{tabular}           & \begin{tabular}[c]{@{}c@{}}\textbf{0.2920}\\ $\pm$ 0.002\end{tabular}           & \begin{tabular}[c]{@{}c@{}}\textbf{0.3596}\\ $\pm$ 0.002\end{tabular}           & \textbf{2.2307*}         & \textbf{0.3949*} & \textbf{1548*}  & \begin{tabular}[c]{@{}c@{}}0.2664*\\ $\pm$ 0.002\end{tabular}          & \begin{tabular}[c]{@{}c@{}}0.2876*\\ $\pm$ 0.001\end{tabular}          & \begin{tabular}[c]{@{}c@{}}0.3564*\\ $\pm$ 0.002\end{tabular}          & \textbf{2.2427*} & \textbf{0.3959*} & \textbf{1519*}  \\
                                & \textit{2}              & \textbf{\begin{tabular}[c]{@{}c@{}}0.2694\\ $\pm$ 0.002\end{tabular}}  & \textbf{\begin{tabular}[c]{@{}c@{}}0.2918\\ $\pm$ 0.002\end{tabular}}  & \textbf{\begin{tabular}[c]{@{}c@{}}\textbf{0.3594}\\ $\pm$ 0.002\end{tabular}}  & \textbf{2.2490*}         & \textbf{0.3960*} & \textbf{1557*}  & \textbf{\begin{tabular}[c]{@{}c@{}}0.2700*\\ $\pm$ 0.001\end{tabular}} & \textbf{\begin{tabular}[c]{@{}c@{}}0.2924\\ $\pm$ 0.002\end{tabular}}  & \begin{tabular}[c]{@{}c@{}}\textbf{0.3594}\\ $\pm$ 0.000\end{tabular}           & \textbf{2.2672*} & 0.3978* & \textbf{1502*}  \\
                                & \textit{5}              & \textbf{\begin{tabular}[c]{@{}c@{}}0.2702*\\ $\pm$ 0.001\end{tabular}} & \textbf{\begin{tabular}[c]{@{}c@{}}0.2934*\\ $\pm$ 0.001\end{tabular}} & \textbf{\begin{tabular}[c]{@{}c@{}}0.3610\\ $\pm$ 0.001\end{tabular}}  & \textbf{2.2063}          & \textbf{0.3939*} & \textbf{1509*}  & \begin{tabular}[c]{@{}c@{}}0.2622*\\ $\pm$ 0.001\end{tabular}          & \begin{tabular}[c]{@{}c@{}}0.2826*\\ $\pm$ 0.001\end{tabular}          & \begin{tabular}[c]{@{}c@{}}0.3506\\ $\pm$ 0.001\end{tabular}           & \textbf{2.2595*} & \textbf{0.3985*} & \textbf{1492*}  \\
                                & \textit{10}             & \begin{tabular}[c]{@{}c@{}}0.2686\\ $\pm$ 0.002\end{tabular}           & \textbf{\begin{tabular}[c]{@{}c@{}}0.2916\\ $\pm$ 0.003\end{tabular}}  & \begin{tabular}[c]{@{}c@{}}0.3586\\ $\pm$ 0.002\end{tabular}           & \textbf{2.2579*}         & \textbf{0.3972*} & \textbf{1554*}  & \begin{tabular}[c]{@{}c@{}}0.2636*\\ $\pm$ 0.002\end{tabular}          & \begin{tabular}[c]{@{}c@{}}0.2864*\\ $\pm$ 0.003\end{tabular}          & \begin{tabular}[c]{@{}c@{}}0.3532\\ $\pm$ 0.003\end{tabular}           & \textbf{2.2221*} & \textbf{0.3950*} & \textbf{1521*}  \\ \hline
\multirow{5}{*}{\textbf{NPA}}   & \textit{0}              & \begin{tabular}[c]{@{}c@{}}\textbf{0.2740}\\ $\pm$ 0.002\end{tabular}           & \begin{tabular}[c]{@{}c@{}}0.2968\\ $\pm$ 0.003\end{tabular}           & \begin{tabular}[c]{@{}c@{}}0.3606\\ $\pm$ 0.002\end{tabular}           & 2.1994                   & \textbf{0.3906}           & 1314            & \begin{tabular}[c]{@{}c@{}}0.2740\\ $\pm$ 0.002\end{tabular}           & \begin{tabular}[c]{@{}c@{}}0.2968\\ $\pm$ 0.003\end{tabular}           & \begin{tabular}[c]{@{}c@{}}0.3606\\ $\pm$ 0.002\end{tabular}           & 2.2103           & 0.3904           & 1328            \\
                                & \textit{1}              & \textbf{\begin{tabular}[c]{@{}c@{}}0.2758\\ $\pm$ 0.003\end{tabular}}  & \textbf{\begin{tabular}[c]{@{}c@{}}0.2984\\ $\pm$ 0.004\end{tabular}}  & \textbf{\begin{tabular}[c]{@{}c@{}}0.3610\\ $\pm$ 0.003\end{tabular}}  & \textbf{2.1975} & 0.3898  & \textbf{1323}   & \textbf{\begin{tabular}[c]{@{}c@{}}0.2759\\ $\pm$ 0.003\end{tabular}}  & \textbf{\begin{tabular}[c]{@{}c@{}}0.2978\\ $\pm$ 0.003\end{tabular}}  & \textbf{\begin{tabular}[c]{@{}c@{}}0.3618\\ $\pm$ 0.003\end{tabular}}  & 2.1938           & \textbf{0.3906}  & 1274            \\
                                & \textit{2}              & \textbf{\begin{tabular}[c]{@{}c@{}}0.2754\\ $\pm$ 0.002\end{tabular}}  & \textbf{\begin{tabular}[c]{@{}c@{}}0.2988\\ $\pm$ 0.003\end{tabular}}  & \textbf{\begin{tabular}[c]{@{}c@{}}0.3624\\ $\pm$ 0.003\end{tabular}}  & 2.1702          & 0.3894  & 1269   & \begin{tabular}[c]{@{}c@{}}0.2740\\ $\pm$ 0.002\end{tabular}           & \textbf{\begin{tabular}[c]{@{}c@{}}0.2996\\ $\pm$ 0.003\end{tabular}}  & \textbf{\begin{tabular}[c]{@{}c@{}}0.3612\\ $\pm$ 0.003\end{tabular}}  & \textbf{2.2675}  & \textbf{0.3956}  & \textbf{1347}   \\
                                & \textit{5}              & \textbf{\begin{tabular}[c]{@{}c@{}}0.2764*\\ $\pm$ 0.002\end{tabular}} & \textbf{\begin{tabular}[c]{@{}c@{}}0.300*\\ $\pm$ 0.003\end{tabular}}  & \textbf{\begin{tabular}[c]{@{}c@{}}0.3634*\\ $\pm$ 0.002\end{tabular}} & 2.1805          & \textbf{0.3906}  & 1282            & \textbf{\begin{tabular}[c]{@{}c@{}}0.2774\\ $\pm$ 0.003\end{tabular}}  & \textbf{\begin{tabular}[c]{@{}c@{}}0.3034*\\ $\pm$ 0.002\end{tabular}} & \textbf{\begin{tabular}[c]{@{}c@{}}0.3662*\\ $\pm$ 0.003\end{tabular}} & 2.1511           & 0.3886           & 1316            \\
                                & \textit{10}             & \textbf{\begin{tabular}[c]{@{}c@{}}0.2776\\ $\pm$ 0.003\end{tabular}}  & \textbf{\begin{tabular}[c]{@{}c@{}}0.3014\\ $\pm$ 0.004\end{tabular}}  & \textbf{\begin{tabular}[c]{@{}c@{}}0.3650\\ $\pm$ 0.004\end{tabular}}  & 2.1878          & 0.3904           & 1304            & \textbf{\begin{tabular}[c]{@{}c@{}}0.2768\\ $\pm$ 0.004\end{tabular}}  & \textbf{\begin{tabular}[c]{@{}c@{}}0.3008\\ $\pm$ 0.004\end{tabular}}  & \textbf{\begin{tabular}[c]{@{}c@{}}0.3634\\ $\pm$ 0.004\end{tabular}}  & \textbf{2.2186}           & \textbf{0.3929}  & 1292            \\ \hline
\multirow{5}{*}{\textbf{LSTUR}} & \textit{0}              & \begin{tabular}[c]{@{}c@{}}0.2968\\ $\pm$ 0.005\end{tabular}           & \begin{tabular}[c]{@{}c@{}}0.3248\\ $\pm$ 0.005\end{tabular}           & \begin{tabular}[c]{@{}c@{}}0.3664\\ $\pm$ 0.004\end{tabular}           & 2.3404                   & 0.3979           & 1480            & \begin{tabular}[c]{@{}c@{}}0.2968\\ $\pm$ 0.005\end{tabular}           & \begin{tabular}[c]{@{}c@{}}0.3248\\ $\pm$ 0.005\end{tabular}           & \begin{tabular}[c]{@{}c@{}}0.3664\\ $\pm$ 0.004\end{tabular}           & 2.2243           & 0.3919           & 1480            \\
                                & \textit{1}              & \begin{tabular}[c]{@{}c@{}}0.2952\\ $\pm$ 0.003\end{tabular}           & \begin{tabular}[c]{@{}c@{}}0.3230\\ $\pm$ 0.004\end{tabular}           & \textbf{\begin{tabular}[c]{@{}c@{}}0.3852\\ $\pm$ 0.003\end{tabular}}  & \textbf{2.3820}          & \textbf{0.3995}  & \textbf{1491*}  & \begin{tabular}[c]{@{}c@{}}0.2900*\\ $\pm$ 0.001\end{tabular}          & \begin{tabular}[c]{@{}c@{}}0.3158*\\ $\pm$ 0.002\end{tabular}          & \textbf{\begin{tabular}[c]{@{}c@{}}0.3800\\ $\pm$ 0.001\end{tabular}}  & \textbf{2.3839*} & \textbf{0.3996}  & \textbf{1523*}  \\
                                & \textit{2}              & \begin{tabular}[c]{@{}c@{}}0.2964\\ $\pm$ 0.005\end{tabular}           & \begin{tabular}[c]{@{}c@{}}0.3244\\ $\pm$ 0.006\end{tabular}           & \textbf{\begin{tabular}[c]{@{}c@{}}0.3872\\ $\pm$ 0.004\end{tabular}}  & \textbf{2.3419*}         & \textbf{0.3978}  & 1455*           & \begin{tabular}[c]{@{}c@{}}0.2862*\\ $\pm$ 0.004\end{tabular}          & \begin{tabular}[c]{@{}c@{}}0.3128*\\ $\pm$ 0.004\end{tabular}          & \textbf{\begin{tabular}[c]{@{}c@{}}0.3764\\ $\pm$ 0.003\end{tabular}}  & \textbf{2.3406}  & \textbf{0.3969}  & \textbf{1504*}  \\
                                & \textit{5}              & \textbf{\begin{tabular}[c]{@{}c@{}}0.2968\\ $\pm$ 0.002\end{tabular}}  & \textbf{\begin{tabular}[c]{@{}c@{}}0.3248\\ $\pm$ 0.003\end{tabular}}  & \textbf{\begin{tabular}[c]{@{}c@{}}0.3872\\ $\pm$ 0.003\end{tabular}}  & \textbf{2.3615}          & \textbf{0.3991}  & 1452*           & \begin{tabular}[c]{@{}c@{}}\textbf{0.2974*}\\ $\pm$ 0.002\end{tabular}          & \begin{tabular}[c]{@{}c@{}}0.3126*\\ $\pm$ 0.002\end{tabular}          & \textbf{\begin{tabular}[c]{@{}c@{}}0.3754\\ $\pm$ 0.001\end{tabular}}  & \textbf{2.2614*} & \textbf{0.3935*} & 1470*           \\
                                & \textit{10}             & \begin{tabular}[c]{@{}c@{}}0.2962\\ $\pm$ 0.005\end{tabular}           & \begin{tabular}[c]{@{}c@{}}0.3236\\ $\pm$ 0.006\end{tabular}           & \textbf{\begin{tabular}[c]{@{}c@{}}0.3864\\ $\pm$ 0.005\end{tabular}}  & \textbf{2.3407}          & \textbf{0.3991}  & 1442*           & \begin{tabular}[c]{@{}c@{}}0.2884\\ $\pm$ 0.005\end{tabular}           & \begin{tabular}[c]{@{}c@{}}0.3138\\ $\pm$ 0.005\end{tabular}           & \textbf{\begin{tabular}[c]{@{}c@{}}0.3766\\ $\pm$ 0.004\end{tabular}}  & \textbf{2.2915*} & \textbf{0.3944}  & 1371*           \\ \hline
\end{tabular}%
}}
\end{table}

On the MIND data, NRMS experiences improvement at $\lambda = $ 2\% and 5\% for the one-step variant, and only at 2\% for the two-step variant.
NPA exhibits improved performance across $\lambda$ values for both variants. 
LSTUR consistantly improves on nDCG@10, but decreases for $\lambda= 1\%, 2\%, 10\%$ on MRR and nDCG@5. 
On the GoodBook data, we observe that all the algorithms outperform MostPop. 
The recommendation performance of ItemKNN and ImplicitMF varies across different levels of $\lambda$, while BPR consistently maintains higher accuracy across $\lambda$ levels, particularly at 1\% and 2\%. 
In book recommendation, performance seems to decrease at the highest $\lambda$ values. This phenomenon may be attributed to the CF algorithms' sensitivity to popularity bias~\cite{deldjoo2021explaining,Savvina2023Reproducing}. A higher $\lambda$ means a larger change in the long-tail distribution of item exposure (in the train set, not in the test set), which negatively impacts the accuracy. 

In general, the observed differences in accuracy are small (maximum increase is 0.002 for GoodBook on BPR; maximum decrease is 0.034 for the same data and algorithm), signifying that performance remains relatively stable when we apply our approach. Many of the observed differences are not statistically significant. In the one-step variant, only some observed increases in accuracy are significant. In the two-step approach, we observe both significant increases and decreases in performance. This is in line with our expectations, that a larger profile (as in the one-step variant) leads to improved performance. We see that algorithms react differently to different $\lambda$ values. This emphasizes the need to carefully select $\lambda$ to maintain recommendation quality.

\begin{table}[htb]
\centering
\caption{Recommendation performance for the GoodBook Data. 
We emphasize the recommendations with the highest scores across conditions (original vs pre-processed data). Statistical significance is indicated by $^*$, based on paired t-tests comparing $\lambda$ values to the baseline ($\lambda = 0$). Standard deviation is denoted by $\pm$.}

\label{tab:Step-preprocess-GoodRead_AccDiv}
\resizebox{\textwidth}{!}{%
\begin{tabular}{cccccccccccccccc}
\cline{3-16}
\multicolumn{2}{c}{}                                          & \multicolumn{7}{c}{One-step Pre-process}                                                                                                                                                                                                                                                  & \multicolumn{7}{c}{Two-step Pre-process}                                                                                                                                                                                                                                                                            \\ \cline{3-16} 
\multicolumn{2}{c}{}                                          & \multicolumn{3}{c}{Accuracy}                                                                                                                                                                                   & \multicolumn{2}{c}{Calibration@10}  & \multicolumn{2}{c}{Descriptive@10} & \multicolumn{3}{c}{Accuracy}                                                                                                                                                                                                             & \multicolumn{2}{c}{Calibration@10}  & \multicolumn{2}{c}{Descriptive@10} \\ \hline
\textit{\textbf{Algorithms}}         & \textit{$lambda$ (\%)} & \textit{MRR}                                                        & \textit{nDCG@5}                                                    & nDCG@10                                                             & \textit{KL}      & \textit{JS}      & \textit{Cov}   & \textit{Gini}     & \textit{MRR}                                                                 & \textit{nDCG@5}                                                             & \textit{nDCG@10}                                                            & \textit{KL}      & \textit{JS}      & \textit{Cov}   & \textit{Gini}     \\ \specialrule{.2em}{.1em}{.1em}
\multirow{5}{*}{\textbf{MostPop}}    & \textit{0}             & \begin{tabular}[c]{@{}c@{}}0.0739\\ $\pm$ 0.00\end{tabular}            & \begin{tabular}[c]{@{}c@{}}0.0347\\ $\pm$ 0.00\end{tabular}           & \begin{tabular}[c]{@{}c@{}}0.0507\\ $\pm$ 0.00\end{tabular}            & 0.0626           & 0.1784           & 31             & 0.3493            & \begin{tabular}[c]{@{}c@{}}0.0739\\ $\pm$ 0.00\end{tabular}                     & \begin{tabular}[c]{@{}c@{}}0.0347\\ $\pm$ 0.00\end{tabular}                    & \begin{tabular}[c]{@{}c@{}}0.0507\\ $\pm$ 0.00\end{tabular}                    & 0.0626           & 0.1784           & 31             & 0.3493            \\
                                     & \textit{1}             & \textbf{\begin{tabular}[c]{@{}c@{}}0.0739*\\ $\pm$ 0.00\end{tabular}}  & \textbf{\begin{tabular}[c]{@{}c@{}}0.0347*\\ $\pm$ 0.00\end{tabular}} & \begin{tabular}[c]{@{}c@{}}0.0507*\\ $\pm$ 0.00\end{tabular}           & \textbf{0.1985*} & \textbf{0.3091*} & \textbf{31}    & \textbf{0.3493}   & \begin{tabular}[c]{@{}c@{}}0.0730*\\ $\pm$ 0.00\end{tabular}                    & \begin{tabular}[c]{@{}c@{}}0.0317*\\ $\pm$ 0.00\end{tabular}                   & \begin{tabular}[c]{@{}c@{}}0.0461*\\ $\pm$ 0.00\end{tabular}                   & \textbf{0.1988*} & \textbf{0.3090*} & 29             & \textbf{0.3707}   \\
                                     & \textit{2}             & \textbf{\begin{tabular}[c]{@{}c@{}}0.0742*\\ $\pm$ 0.00\end{tabular}}  & \textbf{\begin{tabular}[c]{@{}c@{}}0.0348*\\ $\pm$ 0.00\end{tabular}} & \textbf{\begin{tabular}[c]{@{}c@{}}0.0510*\\ $\pm$ 0.00\end{tabular}}  & \textbf{0.1994}  & \textbf{0.3095}  & \textbf{31}    & \textbf{0.3494}   & \begin{tabular}[c]{@{}c@{}}0.0686*\\ $\pm$ 0.00\end{tabular}                    & \begin{tabular}[c]{@{}c@{}}0.0272*\\ $\pm$ 0.00\end{tabular}                   & \begin{tabular}[c]{@{}c@{}}0.0461*\\ $\pm$ 0.00\end{tabular}                   & \textbf{0.1999}  & \textbf{0.3094}  & 29             & \textbf{0.3699}   \\
                                     & \textit{5}             & \begin{tabular}[c]{@{}c@{}}0.0494*\\ $\pm$ 0.00\end{tabular}           & \textbf{\begin{tabular}[c]{@{}c@{}}0.0389*\\ $\pm$ 0.00\end{tabular}} & \begin{tabular}[c]{@{}c@{}}0.0348*\\ $\pm$ 0.00\end{tabular}           & \textbf{0.2005}  & \textbf{0.3102}  & \textbf{31}    & \textbf{0.3499}   & \begin{tabular}[c]{@{}c@{}}0.0438*\\ $\pm$ 0.00\end{tabular}                    & \begin{tabular}[c]{@{}c@{}}0.0172*\\ $\pm$ 0.00\end{tabular}                   & \begin{tabular}[c]{@{}c@{}}0.0298*\\ $\pm$ 0.00\end{tabular}                   & \textbf{0.1809}  & \textbf{0.2976}  & 25             & \textbf{0.4246}   \\
                                     & \textit{10}            & \textbf{\begin{tabular}[c]{@{}c@{}}0.0739*\\ $\pm$ 0.00\end{tabular}}  & \begin{tabular}[c]{@{}c@{}}0.0317*\\ $\pm$ 0.00\end{tabular}          & \begin{tabular}[c]{@{}c@{}}0.0476*\\ $\pm$ 0.00\end{tabular}           & \textbf{0.1968}  & \textbf{0.3029}  & \textbf{31}    & \textbf{0.3501}   & \textbf{\begin{tabular}[c]{@{}c@{}}0.0758*\\ $\pm$ 0.00\end{tabular}}           & \begin{tabular}[c]{@{}c@{}}0.0277*\\ $\pm$ 0.00\end{tabular}                   & \begin{tabular}[c]{@{}c@{}}0.0415*\\ $\pm$ 0.00\end{tabular}                   & \textbf{0.1766}  & \textbf{0.2883}  & 20             & \textbf{0.5227}   \\ \specialrule{.2em}{.1em}{.1em}
\multirow{5}{*}{\textbf{ItemKNN}}    & \textit{0}             & \begin{tabular}[c]{@{}c@{}}0.3881\\ $\pm$ 0.01\end{tabular}            & \begin{tabular}[c]{@{}c@{}}0.1821\\ $\pm$ 0.007\end{tabular}          & \begin{tabular}[c]{@{}c@{}}0.2248\\ $\pm$ 0.005\end{tabular}           & 0.0819           & 0.1944           & 683            & 0.3471            & \begin{tabular}[c]{@{}c@{}}0.3881\\ $\pm$ 0.01\end{tabular}                     & \begin{tabular}[c]{@{}c@{}}0.1821\\ $\pm$ 0.007\end{tabular}                   & \begin{tabular}[c]{@{}c@{}}0.2248\\ $\pm$ 0.005\end{tabular}                   & 0.0819           & 0.1944           & 683            & 0.3471            \\
                                     & \textit{1}             & \begin{tabular}[c]{@{}c@{}}0.3793\\ $\pm$ 0.01\end{tabular}            & \begin{tabular}[c]{@{}c@{}}0.1791\\ $\pm$ 0.002\end{tabular}          & \begin{tabular}[c]{@{}c@{}}0.2203\\ $\pm$ 0.008\end{tabular}           & \textbf{0.1711}  & \textbf{0.2899}  & 665            & \textbf{0.3503}   & \begin{tabular}[c]{@{}c@{}}0.3661\\ $\pm$ 0.002\end{tabular}         & \begin{tabular}[c]{@{}c@{}}0.1756\\ $\pm$ 0.003\end{tabular}        & \begin{tabular}[c]{@{}c@{}}0.2134*\\ $\pm$ 0.002\end{tabular}                  & \textbf{0.1711}  & \textbf{0.2899}  & 665            & \textbf{0.3503}   \\
                                     & \textit{2}             & \textbf{\begin{tabular}[c]{@{}c@{}}0.3894\\ $\pm$ 0.008\end{tabular}}  & \textbf{\begin{tabular}[c]{@{}c@{}}0.1824\\ $\pm$ 0.006\end{tabular}} & \begin{tabular}[c]{@{}c@{}}0.2209\\ $\pm$ 0.004\end{tabular}           & \textbf{0.1712}  & \textbf{0.2892}  & 663            & 0.3345            & \textbf{\begin{tabular}[c]{@{}c@{}}0.3899*\\ $\pm$ 0.006\end{tabular}} & \begin{tabular}[c]{@{}c@{}}0.1713\\ $\pm$ 0.002\end{tabular}          & \textbf{\begin{tabular}[c]{@{}c@{}}0.2294*\\ $\pm$ 0.002\end{tabular}}         & \textbf{0.1719}  & \textbf{0.3094}  & 663            & 0.3384            \\
                                     & \textit{5}             & \begin{tabular}[c]{@{}c@{}}0.3566\\ $\pm$ 0.004\end{tabular}           & \begin{tabular}[c]{@{}c@{}}0.1725*\\ $\pm$ 0.005\end{tabular}         & \begin{tabular}[c]{@{}c@{}}0.2061*\\ $\pm$ 0.002\end{tabular}          & \textbf{0.1741}  & \textbf{0.2906}  & 662            & 0.2886            & \begin{tabular}[c]{@{}c@{}}0.3258*\\ $\pm$ 0.010\end{tabular}                   & \begin{tabular}[c]{@{}c@{}}0.1576*\\ $\pm$ 0.004\end{tabular}                  & \begin{tabular}[c]{@{}c@{}}0.1884*\\ $\pm$ 0.004\end{tabular}                  & \textbf{0.1747}  & \textbf{0.2916}  & 665            & \textbf{0.3503}   \\
                                     & \textit{10}            & \begin{tabular}[c]{@{}c@{}}0.3582*\\ $\pm$ 0.007\end{tabular}          & \begin{tabular}[c]{@{}c@{}}0.1714\\ $\pm$ 0.002\end{tabular}          & \begin{tabular}[c]{@{}c@{}}0.2048*\\ $\pm$ 0.002\end{tabular}          & \textbf{0.1746}  & \textbf{0.2907}  & 602            & 0.2249            & \begin{tabular}[c]{@{}c@{}}0.3378*\\ $\pm$ 0.007\end{tabular}                   & \begin{tabular}[c]{@{}c@{}}0.1576*\\ $\pm$ 0.006\end{tabular}                  & \begin{tabular}[c]{@{}c@{}}0.1949*\\ $\pm$ 0.001\end{tabular}                  & \textbf{0.1729}  & \textbf{0.2905}  & 663            & 0.2961            \\ \specialrule{.2em}{.1em}{.1em}
\multirow{5}{*}{\textbf{ImplicitMF}} & \textit{0}             & \begin{tabular}[c]{@{}c@{}}0.3706\\ $\pm$ 0.003\end{tabular}           & \begin{tabular}[c]{@{}c@{}}0.1793\\ $\pm$ 0.003\end{tabular}          & \begin{tabular}[c]{@{}c@{}}0.2192\\ $\pm$ 0.002\end{tabular}           & 0.0853           & 0.1998           & 608            & 0.4629            & \begin{tabular}[c]{@{}c@{}}0.3706\\ $\pm$ 0.003\end{tabular}                    & \begin{tabular}[c]{@{}c@{}}0.1793\\ $\pm$ 0.003\end{tabular}                   & \begin{tabular}[c]{@{}c@{}}0.2192\\ $\pm$ 0.002\end{tabular}                   & 0.0853           & 0.1998           & 608            & 0.4629            \\
                                     & \textit{1}             & \textbf{\begin{tabular}[c]{@{}c@{}}0.3719*\\ $\pm$ 0.003\end{tabular}} & \textbf{\begin{tabular}[c]{@{}c@{}}0.1804\\ $\pm$ 0.005\end{tabular}} & \textbf{\begin{tabular}[c]{@{}c@{}}0.2213*\\ $\pm$ 0.002\end{tabular}} & \textbf{0.1693}  & \textbf{0.2890}  & 607            & \textbf{0.4661}   & \begin{tabular}[c]{@{}c@{}}0.3533*\\ $\pm$ 0.004\end{tabular}          & \begin{tabular}[c]{@{}c@{}}0.1656*\\ $\pm$ 0.001\end{tabular}         & \begin{tabular}[c]{@{}c@{}}0.2100*\\ $\pm$ 0.001\end{tabular}         & \textbf{0.1701}  & \textbf{0.2892}  & 603            & \textbf{0.4695}   \\
                                     & \textit{2}             & \begin{tabular}[c]{@{}c@{}}0.3659*\\ $\pm$ 0.007\end{tabular}          & \begin{tabular}[c]{@{}c@{}}0.1755\\ $\pm$ 0.002\end{tabular}          & \begin{tabular}[c]{@{}c@{}}0.2152\\ $\pm$ 0.003\end{tabular}           & \textbf{0.1716}  & \textbf{0.2894}  & 605            & \textbf{0.4666}   & \begin{tabular}[c]{@{}c@{}}0.3531*\\ $\pm$ 0.013\end{tabular}          & \begin{tabular}[c]{@{}c@{}}0.1648*\\ $\pm$ 0.00\end{tabular}          & \begin{tabular}[c]{@{}c@{}}0.2027\\ $\pm$ 0.005\end{tabular}          & \textbf{0.1720}  & \textbf{0.2911}  & 605            & \textbf{0.4679}   \\
                                     & \textit{5}             & \begin{tabular}[c]{@{}c@{}}0.3481*\\ $\pm$ 0.001\end{tabular}          & \begin{tabular}[c]{@{}c@{}}0.1656*\\ $\pm$ 0.004\end{tabular}         & \begin{tabular}[c]{@{}c@{}}0.2035*\\ $\pm$ 0.002\end{tabular}          & \textbf{0.1732}  & \textbf{0.2911}  & \textbf{630}   & 0.4533            & \begin{tabular}[c]{@{}c@{}}0.3265*\\ $\pm$ 0.004\end{tabular}                   & \begin{tabular}[c]{@{}c@{}}0.1518*\\ 0.001\end{tabular}                     & \begin{tabular}[c]{@{}c@{}}0.1892*\\ $\pm$ 0.003\end{tabular}                  & \textbf{0.1724}  & \textbf{0.2912}  & \textbf{643}   & 0.4554            \\
                                     & \textit{10}            & \begin{tabular}[c]{@{}c@{}}0.3516*\\ $\pm$ 0.006\end{tabular}          & \begin{tabular}[c]{@{}c@{}}0.1635*\\ $\pm$ 0.004\end{tabular}         & \begin{tabular}[c]{@{}c@{}}0.2042*\\ $\pm$0.001\end{tabular}           & \textbf{0.1724}  & \textbf{0.2895}  & \textbf{649}   & \textbf{0.4654}   & \begin{tabular}[c]{@{}c@{}}0.3346*\\ $\pm$ 0.004\end{tabular}                   & \begin{tabular}[c]{@{}c@{}}0.1546*\\ $\pm$ 0.005\end{tabular}                  & \begin{tabular}[c]{@{}c@{}}0.1931*\\ $\pm$ 0.00\end{tabular}                   & \textbf{0.1696}  & \textbf{0.2875}  & \textbf{657}   & \textbf{0.4889}   \\ \specialrule{.2em}{.1em}{.1em}
\multirow{5}{*}{\textbf{BPR}}        & \textit{0}             & \begin{tabular}[c]{@{}c@{}}0.3469\\ $\pm$ 0.00\end{tabular}            & \begin{tabular}[c]{@{}c@{}}0.1619\\ $\pm$ 0.00\end{tabular}           & \begin{tabular}[c]{@{}c@{}}0.1957\\ $\pm$ 0.00\end{tabular}            & 0.0837           & 0.1979           & 503            & 0.4463            & \begin{tabular}[c]{@{}c@{}}0.3469\\ $\pm$ 0.00\end{tabular}                     & \begin{tabular}[c]{@{}c@{}}0.1619\\ $\pm$ 0.00\end{tabular}                    & \begin{tabular}[c]{@{}c@{}}0.1957\\ $\pm$ 0.00\end{tabular}                    & 0.0837           & 0.1979           & 503            & 0.4463            \\
                                     & \textit{1}             & \textbf{\begin{tabular}[c]{@{}c@{}}0.3470*\\ $\pm$ 0.00\end{tabular}}  & \textbf{\begin{tabular}[c]{@{}c@{}}0.1624*\\ $\pm$ 0.00\end{tabular}} & \begin{tabular}[c]{@{}c@{}}0.1956*\\ $\pm$ 0.00\end{tabular}           & \textbf{0.1686*} & \textbf{0.2885*} & 495            & \textbf{0.4579}   & \textbf{\begin{tabular}[c]{@{}c@{}}0.3472*\\ $\pm$ 0.00\end{tabular}}  & \begin{tabular}[c]{@{}c@{}}0.1614*\\ $\pm$ 0.00\end{tabular}                   & \textbf{\begin{tabular}[c]{@{}c@{}}0.1993*\\ $\pm$ 0.00\end{tabular}} & \textbf{0.1700*} & \textbf{0.2893*} & 499            & \textbf{0.4481}   \\
                                     & \textit{2}             & \textbf{\begin{tabular}[c]{@{}c@{}}0.3509*\\ $\pm$ 0.00\end{tabular}}  & \begin{tabular}[c]{@{}c@{}}0.1607*\\ $\pm$ 0.00\end{tabular}          & \textbf{\begin{tabular}[c]{@{}c@{}}0.1970*\\ $\pm$ 0.00\end{tabular}}  & \textbf{0.1683*} & \textbf{0.2883*} & 495            & \textbf{0.4489}   & \textbf{\begin{tabular}[c]{@{}c@{}}0.3529*\\ $\pm$ 0.00\end{tabular}}           & \textbf{\begin{tabular}[c]{@{}c@{}}0.1630*\\ $\pm$ 0.00\end{tabular}} & \textbf{\begin{tabular}[c]{@{}c@{}}0.2004\\ $\pm$ 0.00\end{tabular}}  & \textbf{0.1656*} & \textbf{0.2884*} & \textbf{509}   & 0.4353            \\
                                     & \textit{5}             & \begin{tabular}[c]{@{}c@{}}0.2910*\\ $\pm$ 0.00\end{tabular}           & \begin{tabular}[c]{@{}c@{}}0.1331*\\ $\pm$ 0.00\end{tabular}          & \begin{tabular}[c]{@{}c@{}}0.1604*\\ $\pm$ 0.00\end{tabular}           & \textbf{0.1666*} & \textbf{0.2868*} & \textbf{538}   & 0.4358            & \begin{tabular}[c]{@{}c@{}}0.2989*\\ $\pm$ 0.00\end{tabular}                    & \begin{tabular}[c]{@{}c@{}}0.1331*\\ $\pm$ 0.00\end{tabular}                   & \begin{tabular}[c]{@{}c@{}}0.1647*\\ $\pm$ 0.00\end{tabular}                   & \textbf{0.1689*} & \textbf{0.2884*} & \textbf{569}   & 0.4264            \\
                                     & \textit{10}            & \begin{tabular}[c]{@{}c@{}}0.3034*\\ $\pm$ 0.00\end{tabular}           & \begin{tabular}[c]{@{}c@{}}0.1405*\\ $\pm$ 0.00\end{tabular}          & \begin{tabular}[c]{@{}c@{}}0.1690*\\ $\pm$ 0.00\end{tabular}           & \textbf{0.1665*} & \textbf{0.2876*} & \textbf{576}   & 0.3987            & \begin{tabular}[c]{@{}c@{}}0.3169*\\ $\pm$ 0.00\end{tabular}                    & \begin{tabular}[c]{@{}c@{}}0.1404*\\ $\pm$ 0.00\end{tabular}                   & \begin{tabular}[c]{@{}c@{}}0.1744*\\ $\pm$ 0.00\end{tabular}                   & \textbf{0.1657}  & \textbf{0.2859*} & \textbf{615}   & 0.4048            \\ \specialrule{.2em}{.1em}{.1em}
\end{tabular}
}
\end{table}

\section{Recommendation Diversity}
Table~\ref{tab:Step-preprocess-accDiv} and Table~\ref{tab:Step-preprocess-GoodRead_AccDiv} present our diversity analysis for news and book recommendations, respectively.
We report calibration as a normative diversity metric and coverage as well as the Gini index as descriptive diversity metrics. 
Focusing on news recommendations, NRMS consistently yields higher calibration and coverage scores for $\lambda = 1\%, 2\%, 5\%, 10\%$ for both 1-step and 2-step variants, as does LSTUR. 
However, for NPA, we observe mixed results, with diversity going up or down depending on $\lambda$. 
For book recommendations, we see positive reactions to diversity from recommender algorithms with varying $\lambda$ values. 
All four algorithms show more calibrated results on preprocessed data compared to the original data, across all $\lambda$ values. 

However, this trend does not hold to the same extent for the descriptive metrics. 
While ImplicitMF and BPR exhibit optimal coverage results for $\lambda = 5\%, 10\%$, ItemKNN performs best at $\lambda = 1\%$. 
We observe discrepancies between the different diversity metrics, where an increase in calibration metrics does not necessarily translate to an increase in descriptive metrics. 

\section{Provider Fairness}
\begin{figure}[htb]
\centering
\begin{subfigure}{.5\textwidth}
  \centering
  \includegraphics[width=.456\textwidth]{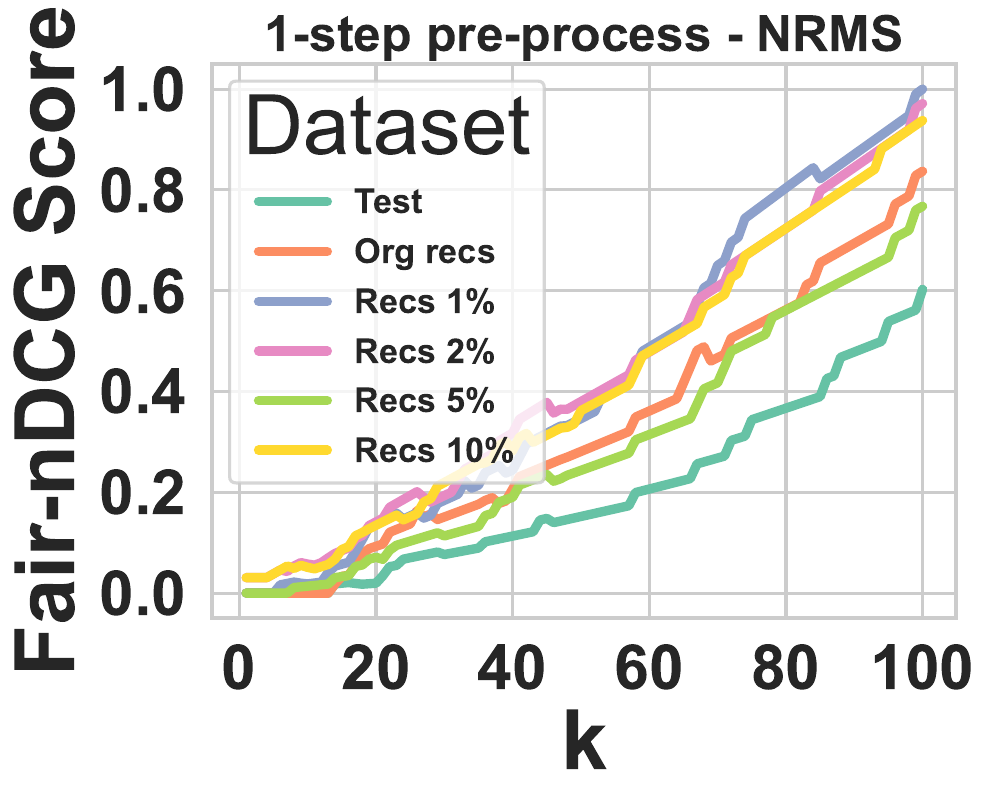}
    \includegraphics[width=.456\textwidth]{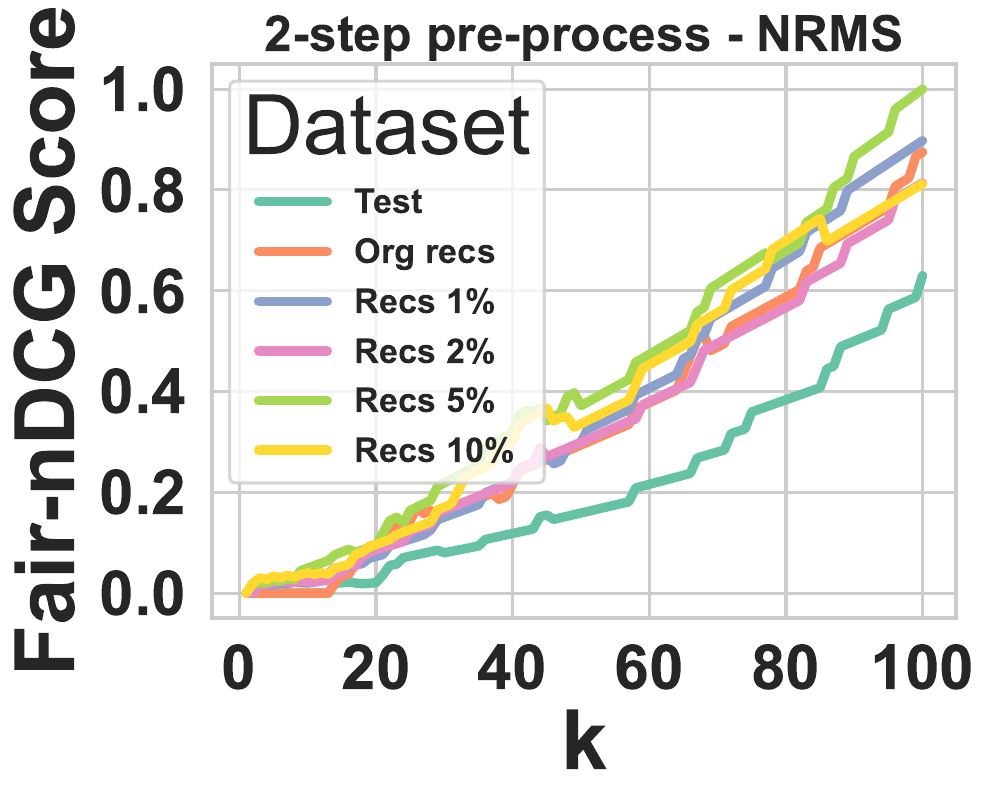}\par
    \includegraphics[width=.456\textwidth]{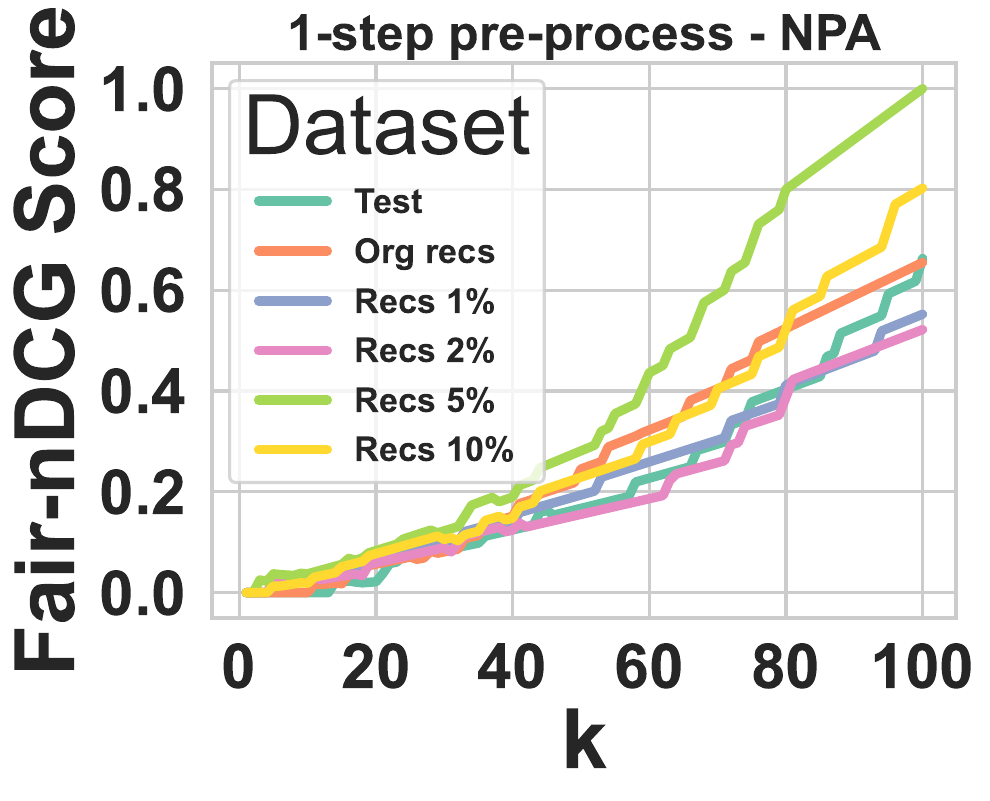}
    \includegraphics[width=.456\textwidth]{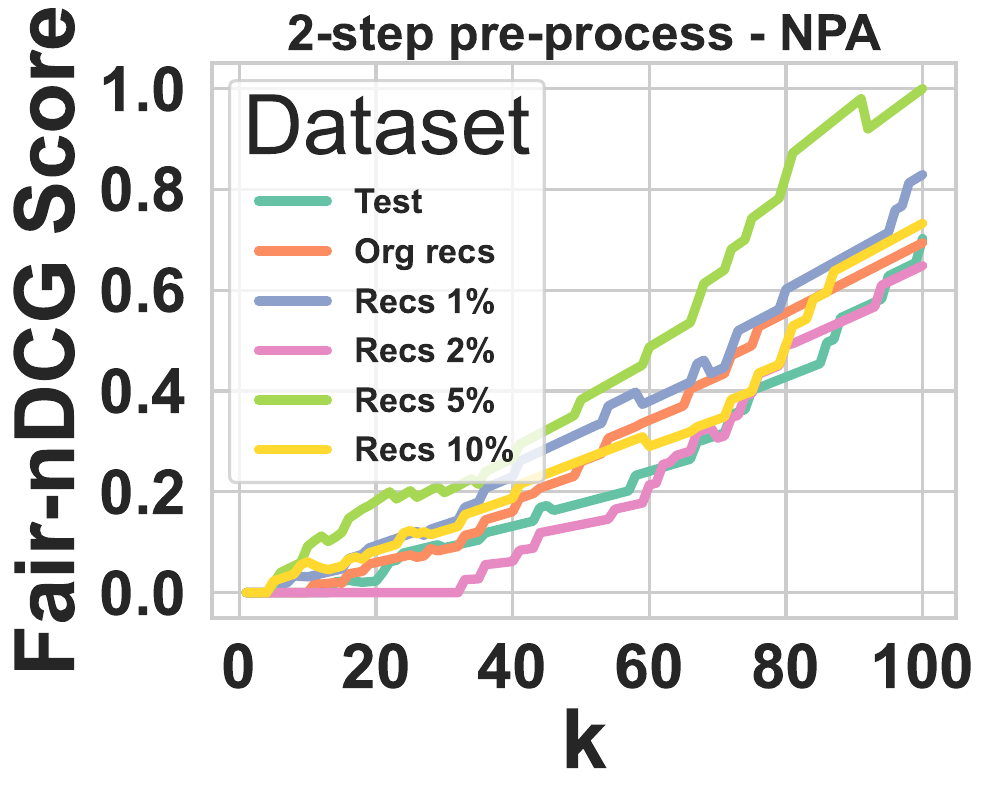}\par
    \includegraphics[width=.456\textwidth]{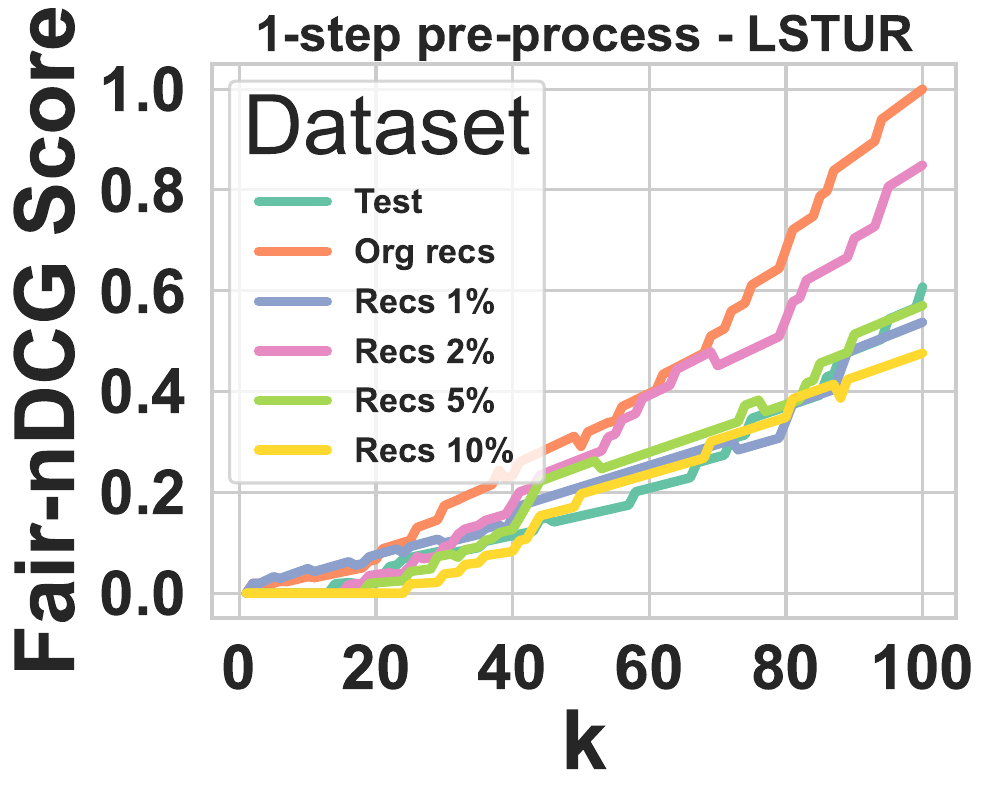}
    \includegraphics[width=.456\textwidth]{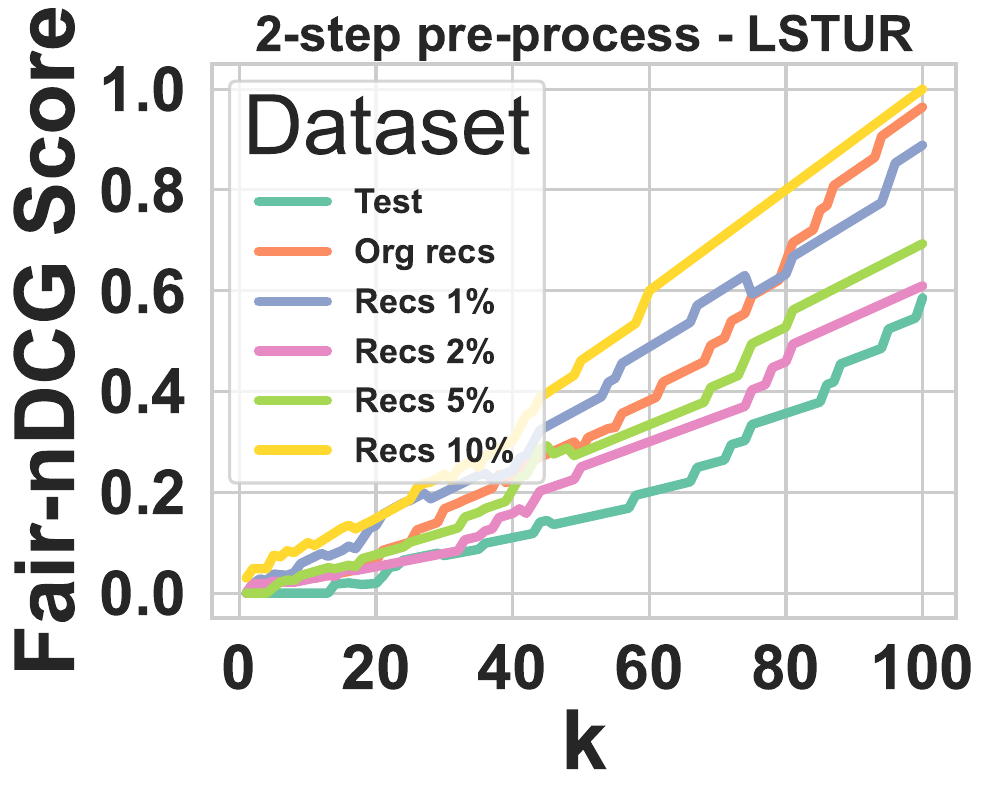}
  \caption{MIND News data}
  \label{fig:fair-ndcg-MIND}
\end{subfigure}%
\begin{subfigure}{.5\textwidth}
  \centering
  \includegraphics[width=.456\textwidth]{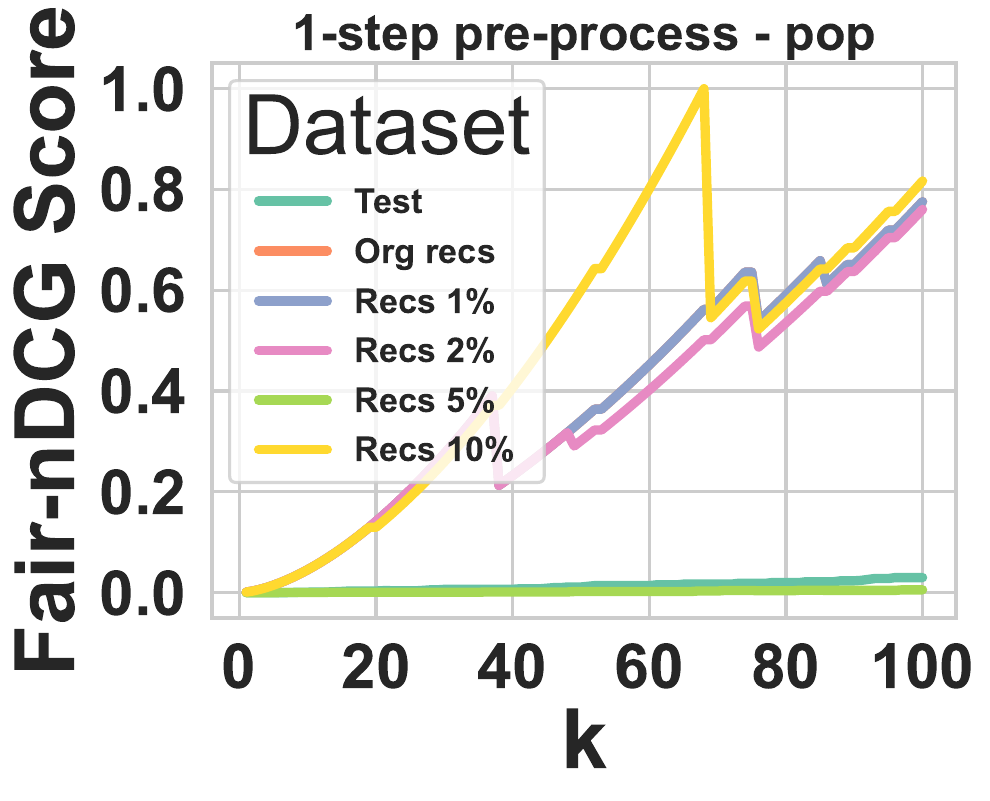}
    \includegraphics[width=.456\textwidth]{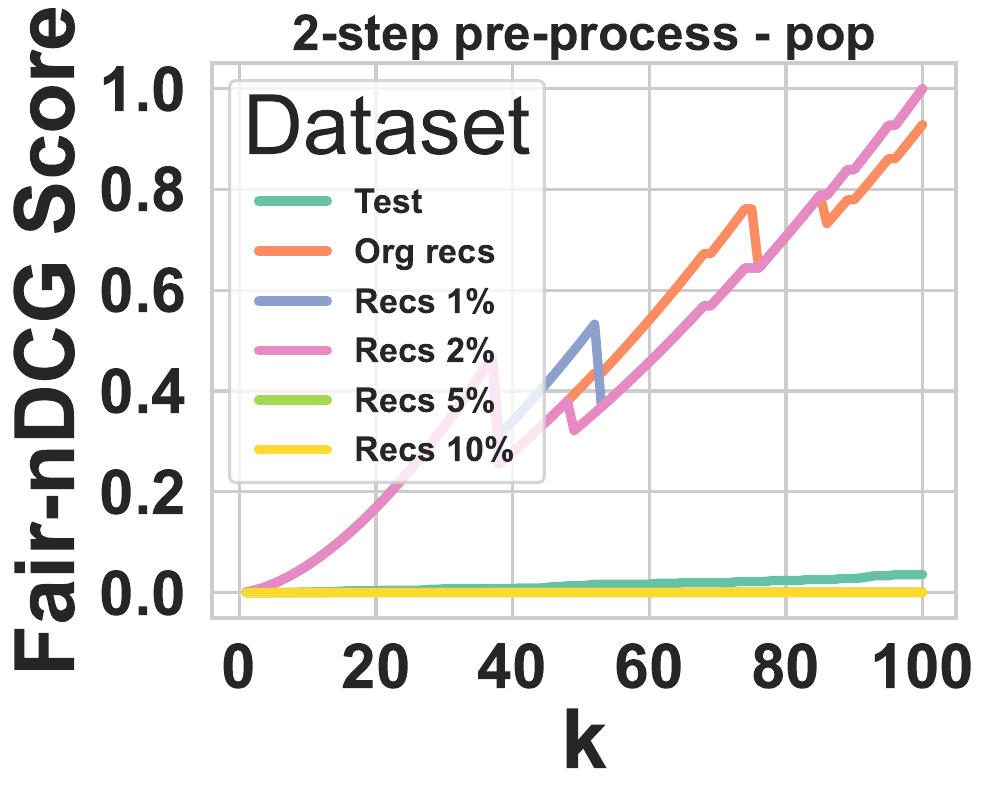}\par
    \includegraphics[width=.456\textwidth]{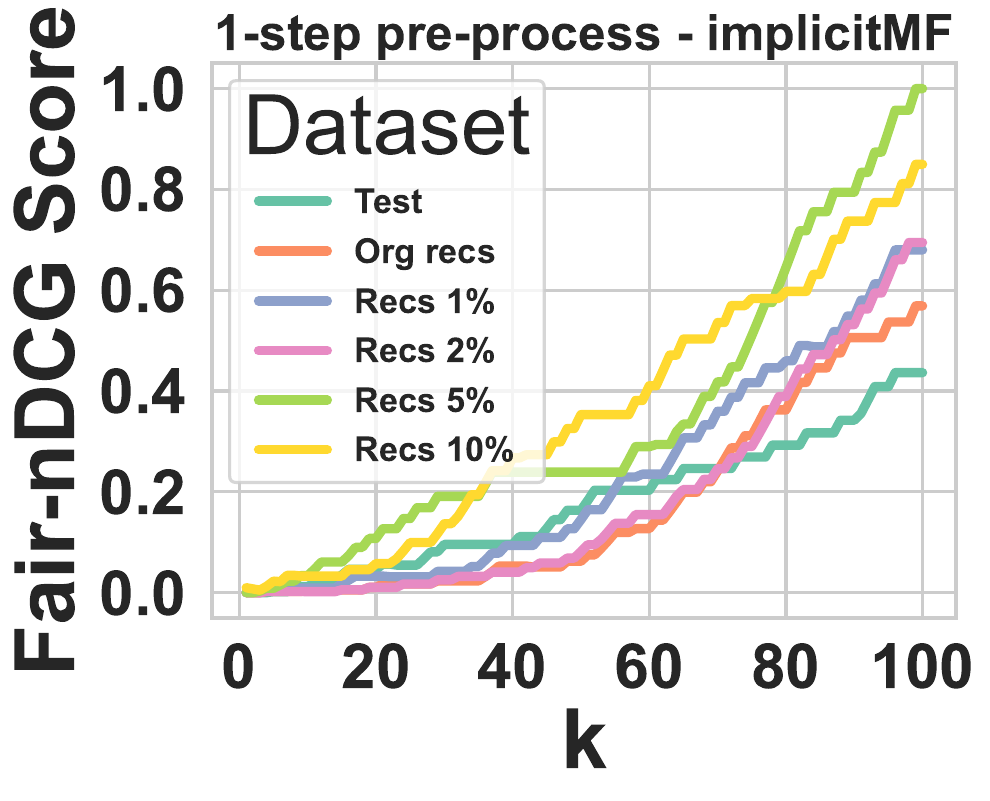}
    \includegraphics[width=.456\textwidth]{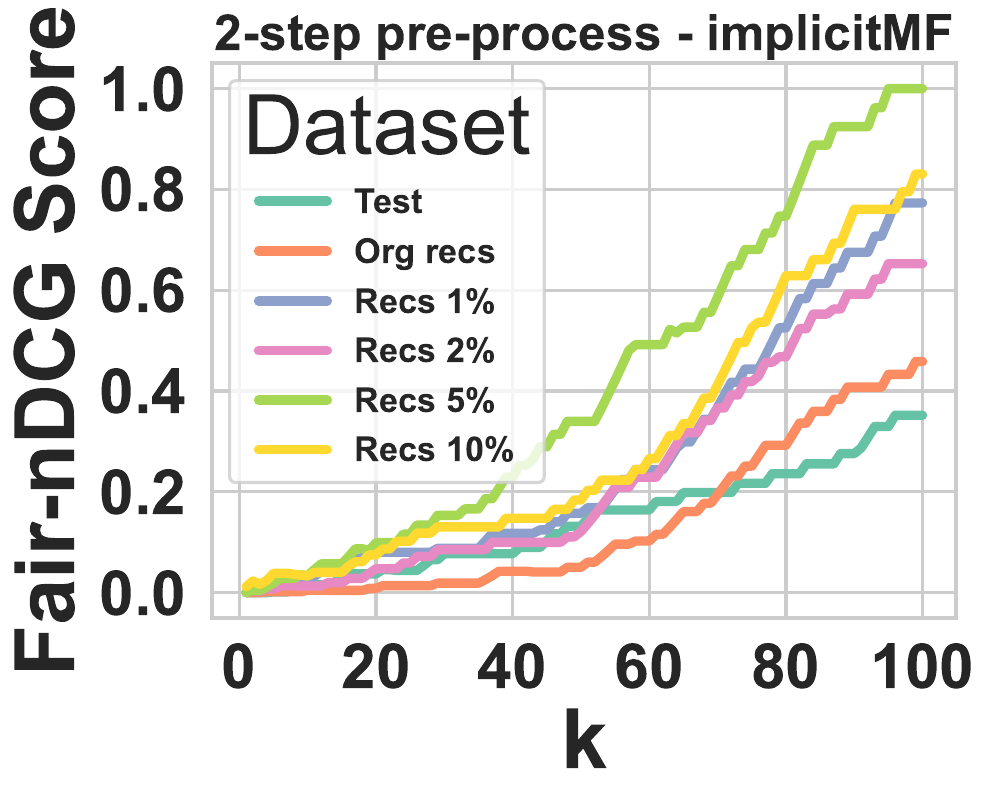}\par
    \includegraphics[width=.456\textwidth]{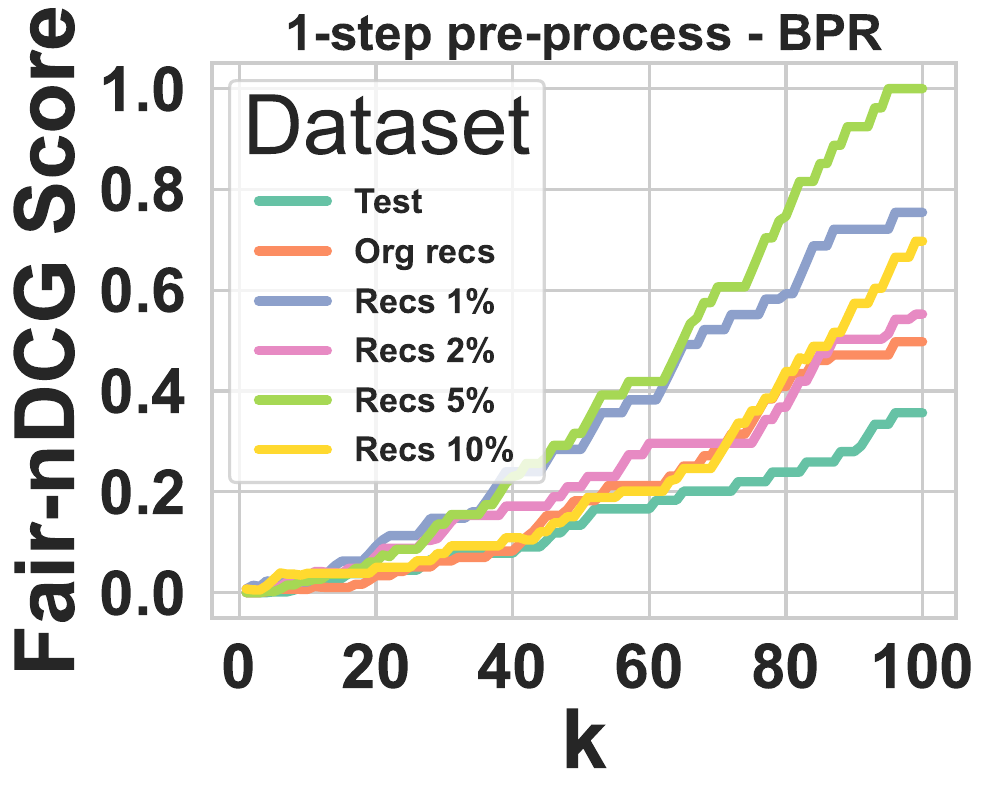}
    \includegraphics[width=.456\textwidth]{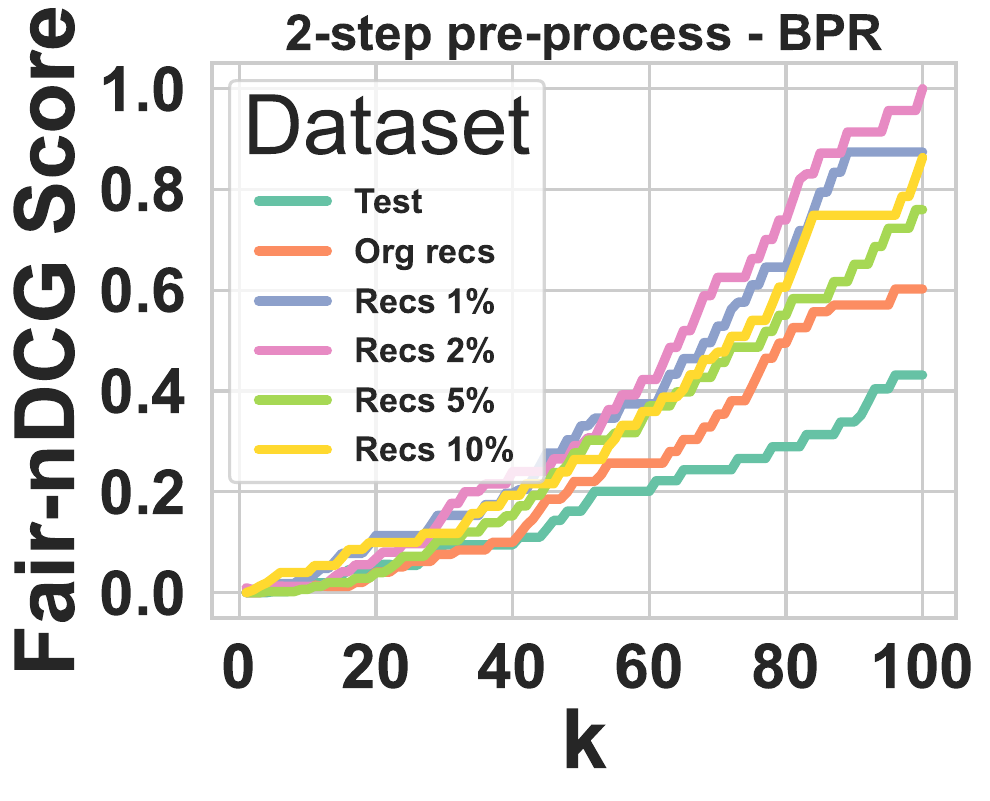}
  \caption{GoodBook Data}
  \label{fig:fair-ndcg-Book}
\end{subfigure}
\caption{Fair-nDCG measured for different top-k recommendation lists. The fair-nDCG-score is measured for the recommendation of the minority \textit{categories}.}
\label{fig:test}
\end{figure}

As mentioned in Section~\ref{sec:intro}, while diversity and provider fairness are based on distinct normative principles, increasing diversity can also contribute to greater provider fairness~\cite{Ekstrand2022}. 
Figure~\ref{fig:test} illustrates 
fair-nDCG results for MIND and GoodBook Data, respectively.  
Each pair of figures corresponds to a specific algorithm, with one figure depicting the results using one-step pre-processing and the other for two-step pre-processing. 
The fair-nDCG scores are evaluated across different levels of $\lambda$ for top-k recommendations ranging from 1 to 100. 
We see that recommendations generated using pre-processed data consistently exhibit higher fair-nDCG scores, indicating increased exposure fairness. 
Also, we observe that the 2-step pre-processing is more effective at lower levels of $\lambda$.  
For instance, in Figure~\ref{fig:fair-ndcg-MIND} (NRMS), recommendations using 10\% one-step variant (and 5\% for two-step) achieve the highest fair-nDCG scores, closely followed by the 1\% and 2\% pre-processing levels.
This emphasizes that pre-processing not only improves the accuracy of the recommendations but also contributes to increased exposure to minority categories. 
For the LSTUR algorithm, we note that at top-k = 20 with 1-step pre-processing, recommendations using the original data ($\lambda = 0$) achieve the best fair-nDCG score. However, with 2-step pre-processing, removing some interactions helps to produce more fair recommendations by increasing the exposure of minority categories ($\lambda = 1\%, 10\%$).
Similar observations hold for GoodBook Data in Figure~\ref{fig:fair-ndcg-Book}. 
In the case of BPR-Figure~\ref{fig:fair-ndcg-Book}, recommendations with 1\% and 5\% One-step variants exhibit the highest fair-nDCG-scores. 
Additionally, recommendations with 2\% and 5\% two-step pre-processing demonstrate noteworthy exposure.

\section{Discussion and Conclusion}
In this paper, we introduced a simple yet effective user-centered pre-processing approach to diversify recommender system outputs. 
The central components of our approach include: (1) using a personalized list of items to maintain accuracy, (2) a well-established selection of categories within user profiles to enhance diversity, and (3) adding and removing $\lambda$ \% of interactions from a user profile.

\paragraph{Main findings} 
Our results demonstrated that keeping data pre-processing closely aligned with user preferences, has the potential to maintain or even improve the performance of the original profiles. 
As for \textit{fairness and diversity}, our experiments have also shown that our pre-processing approach achieves diverse recommendations and simultaneously promotes provider fairness. 
For diversity, through extensive analysis utilizing normative and descriptive diversity measures, we have demonstrated that our approach has the potential to increase item coverage and Gini index and amplify divergence. 
We found that there is a discrepancy between the results of calibration metrics and descriptive metrics for diversity. 
This misalignment could be explained by the fact that normative metrics focus on ideal distributions whereas descriptive metrics capture observed variations in user preferences. 
As a consequence, relying solely on one type of metric could lead to either overestimating or underestimating the true diversity of recommendations.
For fairness, we have demonstrated that our approach actively promotes the recommendation of underrepresented categories. 
This success is driven by (1) using personalized lists to preserve accuracy, (2) strategically selecting user categories to enhance diversity, and (3) adjusting $\lambda$ \% of interactions.

\paragraph{Future work} 
Our paper opens an important new vista for future work. One direction could be on \textit{adapting to changes in user preferences and catalogs}. 
This implies that recommender systems must adaptively and periodically update their data to reflect evolving user preferences and item catalogs. 
Our user-centered pre-processing approach complements the original data by adding extra interactions without altering the core model. 
This dynamic adjustment provides a practical solution to address biases in training data, promoting consistent gains in both fairness and diversity.

Also, another direction could be related to 
\textit{ethical considerations}. 
While our approach aims to improve diversity, it is essential to acknowledge the ethical implications of modifying user profiles~\cite{Grisse2023Manipulation}. 
One key consideration is user autonomy. 
Users have the right to control their data and how it is used. 
Introducing modifications without explicit consent could undermine this autonomy and lead to ethical concerns. 
It is crucial to incorporate mechanisms that allow users to opt in or out of such modifications, giving them control over their data and the recommendations they receive.
Additionally, the explainable nature of our approach opens up new research lines around transparency for end users. 
Future work could investigate to what extent users benefit from knowing how their profile has been altered, or, one step further, how they could benefit from functionality to modify their profiles themselves.  
Another aspect to consider is user autonomy. Users have the right to control their data and how it is used in recommendation systems. 
Introducing modifications to user profiles without their explicit consent could undermine user autonomy and potentially lead to ethical dilemmas. 
It is crucial to incorporate mechanisms that allow users to opt in or opt out of such modifications, giving them greater control over their data and the recommendations they receive.

\section*{Acknowledgment}
This publication is part of the AI, Media \& Democracy Lab (Dutch Research Council project number: NWA.1332.20.009). For more information about the lab and its further activities, visit \url{https://www.aim4dem.nl/}.

%
%
%
\bibliographystyle{splncs04}
\bibliography{sample-base}

\end{document}